\documentclass[aip,reprint]{revtex4-2}
\usepackage[version=4]{mhchem}
\usepackage{dsfont}
\usepackage{booktabs}
\usepackage{graphicx}
\usepackage{xcolor}
\usepackage{hyperref}

\draft % marks overfull lines with a black rule on the right

\newif\ifshowcomments
\showcommentstrue   % \showcommentsfalse to hide

\begin{document}

%\preprint{}

\title{Cooling down trees for finite-temperature quantum dynamics: Purification within ML-MCTDH} %Title of paper



\author{Niclas Krupp}
\email{niclas.krupp@pci.uni-heidelberg.de}
\affiliation{%
 Theoretische Chemie, Physikalisch-Chemisches Institut, Universität Heidelberg, INF 229, 69120 Heidelberg, Germany 
}%

\author{Oriol Vendrell}
\email{oriol.vendrell@uni-heidelberg.de}
\affiliation{%
 Theoretische Chemie, Physikalisch-Chemisches Institut, Universität Heidelberg, INF 229, 69120 Heidelberg, Germany 
}%

\date{\today}

\begin{abstract}
Simulating multidimensional quantum systems at finite temperature is inherently
    challenging as the system is no longer described by a single, pure-state
    wavefunction but by a density operator, squaring an already exponential
    scaling of the Hilbert space.
    Building upon the compact wavefunction ansatz of the multi-layer
    multiconfiguration time-dependent Hartree (ML-MCTDH) method, we present a
    new scheme for simulating finite-temperature quantum dynamics based on
    purification. Here, a density operator is mapped to a single ML-MCTDH
    wavefunction in an enlarged Hilbert space, comprising physical and auxiliary
    degrees of freedom.
    In the key step, one obtains a pure-state representation of the canonical
    density operator via imaginary time-propagation of the infinite-temperature
    state. The most important observation is that this state can be exactly
    decomposed as a Hartree product of maximally entangled combined modes, each
    combined mode consisting of a physical degree of freedom and its auxiliary
    counterpart. Through dynamically pruning the node ranks of the ML-tree
    during the ``cool down'' stage yields a compact finite-temperature
    wavefunction, thus accelerating the real-time propagation and
    enabling finite-temperature simulations of multidimensional, correlated
    molecular systems.
    Our method circumvents both intensive statistical sampling and costly
    tensor-decomposition of the full density-operator, while being broadly
    applicable to model Hamiltonians and general \emph{ab initio} potential energy
    surfaces alike.
    Two applications of the method are presented, benchmark results on the
    thermal ground-state of \ce{H2O}, as well as temperature-dependent infrared
    absorption spectra of the more challenging, floppy \ce{H3O2-} anion.

\end{abstract}


\maketitle %\maketitle must follow title, authors, abstract and \pacs

\section{Introduction}

The interplay of quantum and thermal effects is central to the behavior of
molecules, materials, and biological systems. Even at room temperature and
below, thermal excitation of low-energy vibrational modes, which are ubiquitous
in hydrogen-bonded networks, proteins, and crystal lattices, significantly
influences quantum processes such as
tunneling\cite{tuckerman1997quantum,hernandez2005temperature}, coherent energy
transfer\cite{romero2014quantum,scholes2017using,perlik2014distinguishing}, and
superconductivity\cite{orenstein2000advances,simon1997superconductivity}.
Consequently, simulating the dynamics of quantum systems at finite temperatures
is necessary for understanding biological function and developing quantum
technologies that operate outside the cryogenic regime, among
others\cite{duan2017nature,engel2007evidence,cao2020quantum,scholes2011lessons}.

By finite-temperature quantum dynamics, we mean that the quantum system has
reached thermal equilibrium prior to the simulated dynamics through contact with
a bath. This interaction of a quantum system with an environment produces a
mixture of multiple (normalized but not necessarily orthogonal) pure states
$\lbrace |\phi_i\rangle\rbrace$ such that the system must be described by the
density operator
\begin{gather}
    \hat \rho (t)= \sum_{i}p_i|\phi_i(t)\rangle\langle\phi_i(t)|.
    \label{eq:dens-mat}
\end{gather}
Here, $p_i$ is the probability of finding the system in the state
$|\phi_i\rangle$. We assume that at $t=0$ the system has already undergone
thermalization to temperature $T$, such that the system is initially described
by the canonical density operator,
\begin{gather}
    \hat{\rho}_{\beta}=e^{-\beta\hat H}/Z_\beta,
    \label{eq:botzmann-op}
\end{gather}
with inverse temperature $\beta =1/k_bT$ and partition function $Z_\beta
=\mathrm{Tr}[e^{-\beta\hat H}]$.  As opposed to open quantum system dynamics,
there is no contact with a bath during the real time-evolution of the system.
The resulting unitary time-evolution of the density operator is governed by the
Liouville-von-Neumann (LvN) equation.\cite{may2023charge}

Generally, propagating mixed states is numerically more demanding than
propagating pure states. If the $|\phi_i(t)\rangle$ in Eq.~\eqref{eq:dens-mat}
can be evolved efficiently, the cost of propagating the corresponding density
operator grows only linearly with the number of populated states. In a thermal
mixture, however, this number increases rapidly with temperature, and the
eigenstates of the system are seldom accessible for high-dimensional
Hamiltonians.  Hence, the full density operator becomes, in one form or another,
the starting point for any workable scheme even though its size scales
quadratically with that of the underlying pure-state representation.

Meanwhile, several methods for propagating high-dimensional wavefunctions that
efficiently tackle the underlying exponential scaling were developed over the
past decades. Among them, the multiconfiguration time-dependent Hartree method
(MCTDH) and its multilayer generalization (ML-MCTDH)
\cite{manthe2008multilayer,wang2003multilayer,vendrell2011multilayer} are among the most established and broadly applicable approaches. By mitigating the
exponential scaling through variationally-optimal tensor-decomposition, dynamics
of complex quantum systems with many DOFs become accessible, including highly
flexible water clusters\cite{vendrell2009strong,schroder2022coupling,men26:1}, large
spin-boson models for charge and energy
transport\cite{wang2011numerically,wang2008coherent},
thermal rate constants of bimolecular reactions\cite{matzkies1998accurate,huarte2000full},
as well as molecules strongly coupled to
cavities\cite{mellini2025competition,wallner2024strong,krupp2025quantum}.
Building on the efficiency
of the (ML-)MCTDH method, research has increasingly focused on extending its
utility to treat finite-temperature systems. These efforts can be roughly
divided into three groups:

(\textit{i}) \emph{Tensor-decomposition of the density operator}: Density
operators can be represented as vectors in Liouville space. This space has the square of 
the dimension of the pure-state's Hilbert space, and forms a Hilbert space
itself under the Hilbert–Schmidt inner product. Consequently, tensor
decomposition techniques developed for wavefunctions apply equally to density
operators. In the MCTDH and ML-MCTDH framework, various hierarchical
decompositions of the density matrix in tree-tensor format are possible,
referred to as type I and type II (ML-)MCTDH density
operators\cite{raab_multiconfiguration_1999,raab_multiconfigurational_2000,van_haeften_propagating_2023}.

(\textit{ii}) \emph{Statistical averaging of ensemble wavefunctions}:
$N_{\mathrm{samp}}$ wavefunctions are sampled from the system's Hilbert space,
individually propagated in real-time, and averaged to approximate $\hat\rho(t)$
in Eq.~\eqref{eq:dens-mat}.  To this end, several sampling strategies have been
developed, based, e.g., on random-phase
wavefunctions\cite{gelman_simulating_2003,nest_quantum_2007} or Monte-Carlo
sampling\cite{wang_calculation_2006,wang_quantum-mechanical_2006}.
The total effort corresponds to $N_{\mathrm{samp}}$ times the cost of a
single propagation, which is hopefully smaller than the propagation of the full
density operator, and does not require any new implementation of LvN-related
equations of motion. However, one needs to ensure statistical convergence with the
number of propagated wavepackets, and thus the obtained result is bound by
a statistical uncertainty.

(\textit{iii}) \emph{Purification methods}: Here, the original system's Hilbert
space, referred to as the physical system, is enlarged by a set of artificial
DOFs, commonly termed ancilla or auxiliary system. The density operator is
mapped on a pure state wavefunction within the enlarged Hilbert space where the
time-dependent Schrödinger equation is solved for a pure state wavefunction.
Although, at first sight, purification methods may appear like a mere
reformulation of the direct propagation of the density operator with a similar
scaling cost, they offer greater flexibility and, similarly to statistical
approaches, they do not require the implementation of new equations of motion in
existing codes.
For these reasons, in recent years, thermofield (TF)
dynamics\cite{barnett_liouville_1987,borrelli_quantum_2016}, a special case of
purification methods, has gained attention within the MCTDH community, resulting
in a number of applications of
TF-ML-MCTDH\cite{fischer_thermofield-based_2021,brey_quantum_2021}.

Almost simultaneously, similar strategies were pursued within the condensed
matter physics community and implemented within density matrix renormalization
group (DMRG) or matrix product state (MPS) techniques and codes. This has
resulted in (\textit{i}) the development of matrix product density operators
(MPDOs) for closed and open quantum system
dynamics\cite{verstraete_matrix_2004,zwolak2004mixed}, (\textit{ii}) various
sampling schemes which average over many pure-state MPS
wavefunctions\cite{white_minimally_2009,chen_hybrid_2020}, and (\textit{iii})
application of the ancilla and TF approach to MPS
wavefunctions\cite{feiguin_finite-temperature_2005,verstraete2004density}.

Here, we present an alternative approach to finite-temperature quantum dynamics
based on combining ML-MCTDH and purification.
Besides TF dynamics, purification approaches remain largely unexplored within the
ML-MCTDH methodology; in contrast, in DMRG, such methods have enjoyed a larger
popularity~\cite{feiguin_finite-temperature_2005,verstraete2004density},
although they are limited by the less flexible MPS ansatz compared to ML-MCTDH.

Our approach starts from a maximally compact (to be properly defined below)
representation of an infinite-temperature mixed state cast as a pure state in the
enlarged Hilbert space.
This state is ``cooled down'' by imaginary-time propagation to the desired final
temperature, while pruning configurations below a predefined occupation
threshold in the combined
physical-auxiliary space. Interestingly, our strategy yields the pure ground state
of the system in the $T\to 0$ limit as the only surviving vector.

The paper is organized as follows: First, the concept of purification is briefly
reviewed and the connection to TF dynamics established. Second, we explain how
compact tree-tensor network representations of both the infinite-temperature
and finite-temperature state are achieved within ML-MCTDH by appropriately
introducing auxiliary modes to the ML-tree.
Subsequently, two applications of the method are presented, a study on the
thermal ground-state of the small three-dimensional \ce{H2O} system and the more
challenging, floppy \ce{H3O2-} cluster anion with 9 DOFs. A wide range of
thermal observables can be accessed efficiently from a time-evolved thermal
ML-MCTDH wavefunction, for example temperature-dependent infrared absorption
spectra, and the thermal density of states. We also discuss the
advantages of this approach in relation to existing methods, and
provide perspectives for future extensions of
purification within the ML-MCTDH framework.

\section{Theory}

\subsection{Purification}\label{sec:pur}

In purification, one seeks a pure-state representation of the canonical density
operator at inverse temperature $\beta$, as defined in
Eq.~\eqref{eq:botzmann-op}, where $\hat H$ denotes the Hamilton operator of the
physical system. Specifically, this pure-state representation must enable
the calculation of statistical quantities such as the partition function
$Z_{\beta}$, thermal expectation values $\langle A\rangle_{\beta}$, and thermal
correlation functions $C_{AB}(t)$, given by
\begin{align}
   Z_{\beta} &= \mathrm{Tr}[e^{-\beta\hat{H}}], \label{eq:th-par} \\
\langle A\rangle_{\beta} &= \mathrm{Tr}[\hat\rho_{\beta}\hat A(t)],  \label{eq:th-expect} \\
C_{AB}^{\beta}(t) &=
   \langle A(t)B(0)\rangle_{\beta}= \mathrm{Tr}[\hat\rho_{\beta}\hat A(t)\hat B(0)]\label{eq:th-CAB}
\end{align}
in terms of the density operator, and where
the trace is taken over a complete basis spanning the system's Hilbert space
(we use $\hbar=1$ throughout).
One can achieve a pure-state representation of mixed states by extending the physical
Hilbert space with an auxiliary space of equal dimension.
In the following, operators acting on the physical space are denoted by the usual
hat $(\hat A)$, operators acting on the auxiliary space carry a tilde $(\tilde A)$,
 and operators acting on
combined physical and auxiliary space carry a bar $(\bar A)$.
One introduces now the state in the extended Hilbert space
\begin{gather}
    |\mathds{1}\rangle =\sum_{i=1}^{N}
|\varphi_i\rangle|\tilde \alpha_i\rangle,\label{eq:one-pur}
\end{gather}
where the normalized states $\lbrace|\varphi_i\rangle\rbrace$ span the system's
Hilbert space and each of which is associated with an orthonormal state in the
auxiliary (or tilde) Hilbert space, $\langle \tilde\alpha_i|\tilde\alpha_j\rangle = \delta_{ij}$.
Besides normalization, we have not yet set further conditions
on $|\varphi_i\rangle$. Let us define the purified thermal state (PTS)

\begin{align}
    |\psi_{\beta}\rangle & = Z_{\beta}^{-1/2} e^{-\beta \hat H /2}
                          |\mathds{1}\rangle  \label{eq:map-dens-th} \\
                         & = Z_{\beta}^{-1/2}
                         \sum_{i=1}^{N} \left(
                            e^{-\beta\hat H/2} |\varphi_i\rangle
                        \right) |\tilde\alpha_i\rangle,
\end{align}
recalling that ``hat'' operators act exclusively on the physical space,
and noting the inverse temperature $\beta/2$ in the statistical operator.
We examine now the expectation value of an operator with the PTS,
\begin{equation}
    \label{eq:expval-phi}
    \langle \psi_\beta |\hat A | \psi_\beta \rangle =
    \frac{1}{Z_\beta}\sum_{i=1}^{N}
    \langle \varphi_i | e^{-\beta\hat H/2} \hat A e^{-\beta\hat H/2} |\varphi_i \rangle.
\end{equation}
For this, we expand the still undefined states $|\varphi_i \rangle$ as
\begin{equation}
    \label{eq:trafo}
    |\varphi_i\rangle = \sum_{j=1}^{N_H} a_{ij} |\chi_j\rangle,
\end{equation}
in a complete and orthonormal basis $\{|\chi_j\rangle\}$ spanning
the physical Hilbert space, with $N_H$ being its total dimension
(in practice finite and determined by the choice of primitive basis).
This results in
\begin{equation}
    \label{eq:expval-chi}
    \langle \psi_\beta |\hat A | \psi_\beta \rangle =
    \frac{1}{Z_\beta}\sum_{i=1}^{N}
    \sum_{j,l=1}^{N_H} a_{ij}^* a_{il}
    \langle \chi_j | e^{-\beta\hat H/2} \hat A e^{-\beta\hat H/2} |\chi_l \rangle.
\end{equation}
In order for the expectation value~\eqref{eq:expval-chi} to function as the
trace in Eq.~\eqref{eq:th-expect}, the expansion coefficients $a_{ij}$ must
fulfill $\sum_i^N a_{ij}^* a_{il}= \delta_{jl}$, i.e.,  $a_{ij}$ must be a
unitary $N_H\times N_H$ matrix. Hence, the $|\varphi_i\rangle$ in the
state $|\mathds{1}\rangle$ are any orthonormal and complete basis, and Eq.~\eqref{eq:trafo} is
a unitary transformation between any two such bases.
The normalization of $|\psi_\beta\rangle$ to 1 at all $\beta$ arises from
the factor $Z_\beta^{-1/2}$ in Eq.~\eqref{eq:map-dens-th}.
Finally, Eq.~\eqref{eq:one-pur} becomes
\begin{gather}
    |\mathds{1}\rangle =\sum_{i=1}^{N_H}
|\chi_i\rangle|\tilde\alpha_i\rangle.\label{eq:one-pur-2}
\end{gather}
$|\mathds{1}\rangle$ covers all states of the complete physical Hilbert space and its
defining property is that
all states in any chosen orthonormal basis carry the same weight. This
corresponds to an infinite-temperature state (ITS).

For real-time propagation, the PTS can be evolved under the Hamiltonian
acting \emph{only} on the physical system DOFs, according to the
usual time-dependent Schrödinger equation
\begin{gather}
    i|\dot\psi_{\beta}(t)\rangle = \hat H |\psi_{\beta}(t)\rangle \label{eq:tdse-pur}
\end{gather}
with formal solution $|\psi_{\beta}(t)\rangle=e^{-i\hat Ht}|\psi_{\beta}(0)\rangle$.
Under this prescription, a time-dependent expectation value is readily computed
as
\begin{align}
    \langle A\rangle_\beta(t) &=
    \langle \psi_{\beta}(t)|\hat A|\psi_{\beta}(t)\rangle \\
    &= Z_{\beta}^{-1}\sum_{j=1}^{N_H}
    \langle \chi_j|e^{-\beta \hat H/2}e^{i \hat H t}\hat A e^{-i \hat H t} e^{-\beta\hat H/2}|\chi_j\rangle\label{eq:expr-pur-exp0} \\
    &= \mathrm{Tr}[\rho_{\beta}^{1/2}\hat A(t) \rho_{\beta}^{1/2}].\label{eq:expr-pur-expect}
\end{align}
It becomes clear that the orthonormal auxiliary basis solely acts as a
``tracer'', i.e., it affords the correct trace operation by eliminating
coherent contributions (for which $i\neq j$).

Analogously, thermal correlation functions [Eq.~\eqref{eq:th-CAB}]
are obtained from time-dependent overlaps of purified thermal wavefunctions,
\begin{align}
C_{AB}^{\beta}(t)=
    \langle \psi_{\beta}(t)|\hat{A}|\psi_{\beta}^{B}(t)\rangle,
    \label{eq:th-CAB-pu}
\end{align}
with $|\psi_{\beta}^{B}(0) \rangle = \hat{B}|\psi_{\beta}\rangle$.
For example, the linear absorption spectrum at finite temperature is
obtained from
\begin{gather}
    I_\beta(\omega)\propto \omega\,\mathrm{Re}\,
    \int_0^\infty e^{i\omega t} C_{\mu\mu}^{\beta}(t) dt, \\
    C_{\mu\mu}^{\beta}(t)= \langle \psi_{\beta}(t)|\hat{\mu}|\psi_{\beta}^{\mu}(t)\rangle.
    \label{eq:th-mu-mu}
\end{gather}
The purified state $|\psi_\beta(0)\rangle$ plays here exactly the same role as
the absolute ground state in 0~K simulations, with the difference that the
non-dipole operated initial state needs to be propagated in real time as well
(at  0~K its time-dependent phase factorizes and can be ignored\cite{tannor2008introduction}).

It is worth mentioning that the density operator of the physical system can be
recovered by tracing the density operator of the PTS over the 
auxiliary system, i.e.,
$\hat{\rho}(t) =
    \sum_i  \langle \tilde\alpha_i|\psi_{\beta}(t)\rangle
             \langle \psi_{\beta}(t)|\tilde\alpha_i\rangle$.
Constructing $\hat{\rho}(t)$ is generally avoided though, as it 
is memory-intensive and, in this formulation, an unnecessary quantity; all relevant
observables can be obtained directly from the purified thermal wavefunction.

Finally, since the auxiliary system acts solely as a tracer, any unitary
evolution within the auxiliary space is possible while leaving the purification
in Eq.~\eqref{eq:map-dens-th} invariant under such transformation.  In
particular, denoting the total Hamiltonian of the physical and auxiliary systems
as
\begin{equation}
    \label{eq:tot-H}
    \bar H = \hat H + \tilde h,
\end{equation}
we choose the simple gauge $\tilde h = \tilde{\mathbf{1}}$ [compare with
Eq.~\eqref{eq:tdse-pur}].  This avoids propagation in the auxiliary space, which
can become computationally costly for high-dimensional systems, and is in
contrast to the gauge choice in TF dynamics, shortly discussed below for
completeness.

\subsection{Connection to Thermofield Dynamics}

TF dynamics
\cite{ari85:429,barnett_liouville_1987}
is a special case of purification with the particularity
that the auxiliary system is defined as an exact copy of the physical system.
Specifically, the PTS at inverse temperature $\beta$ takes the form
\begin{equation}
    \label{eq:one-tf}
    |\psi_\beta\rangle = \frac{1}{Z_\beta^{1/2}}
    \sum_{j} e^{-\beta\hat H/2}|\chi_j\rangle |\tilde\chi_j\rangle
\end{equation}
where $|\tilde\chi_j\rangle$ are copies of the basis functions in the
physical system.
The total Hamiltonian for real time evolution
takes now the form
$\bar H = \hat H - \tilde H$ under the Schrödinger equation
\begin{gather}
    i|\dot\psi_{\beta}(t)\rangle = \bar H |\psi_{\beta}(t)\rangle,
    \label{eq:tdse-tf}
\end{gather}
i.e., the
basis states of the auxiliary system propagate backwards in time
under a copy of the system's Hamiltonian.

The basic advantage of the TF formulation is that, for systems consisting of
uncoupled oscillators (bosonic baths) or mean-field Hamiltonians, it is possible
to find a compact unitary (often referred to in this context as a Bogoliubov) transformation
from the ground state of the combined system to the PTS~\cite{bar85:467,%
barnett_liouville_1987,%
harsha_thermofield_2019,%
borrelli_finite_2021},
\begin{equation}
    \label{eq:tf-trafo}
    |\psi_\beta\rangle = e^{-i \bar G } |0\rangle |\tilde 0\rangle,
\end{equation}
thus avoiding the direct (often numerical) application of the statistical
operator as in Eq.~\eqref{eq:one-tf}. In fact, this transformation is usually applied to the Hamiltonian instead of the wavefunction, shifting the temperature dependence to the Hamiltonian\cite{borrelli_quantum_2016,borrelli_finite_2021}.

Moreover, an exact mapping between time-evolution under
the LvN and Schrödinger equations in the physical and enlarged spaces,
respectively, arises under the TF formulation~\cite{barnett_liouville_1987,shu19:134107}
\begin{equation}
    \label{eq:map-tf-lvn}
    i \dot{\hat{\rho}}_\beta = [\hat H, \hat\rho_\beta]
    \Leftrightarrow
    i |\dot\psi_\rho\rangle = \bar H |\psi_\rho\rangle,
\end{equation}
where $|\psi_\rho\rangle = \hat\rho_\beta|\mathds{1}\rangle$.

TF dynamics has found applicability in finite-temperature quantum
chemistry theories~\cite{harsha_thermofield_2019,shu19:134107},
as well as in the simulation of vibronic Hamiltonians using the
ML-MCTDH approach by Fischer
and Saalfrank~\cite{fischer_thermofield-based_2021}.
In the latter application, an analytical transformation
of the form \eqref{eq:tf-trafo} was applied to the uncoupled
normal modes of the molecule, leaving the electronic degree of
freedom unthermalized due to its higher excitation energy. 

In molecular applications, however, no general analytical transformation exists
for arbitrary multidimensional
potentials, e.g., anharmonic \emph{ab initio}
potential energy surfaces, which strictly limits TF to
linear-vibronic coupling Hamiltonians~\cite{fischer_thermofield-based_2021}
and related models. 
Moreover, the real-time evolution in the TF prescription propagates both physical and
auxiliary DOFs, potentially increasing the computational cost for
strongly coupled, multidimensional
Hamiltonians~\cite{schroder2022coupling}. Instead, we have discussed above how
the propagation of the auxiliary system is not required when computing
expectation values and correlation functions, and how it merely constitutes
a gauge freedom in the real time evolution.
Thus, in the following, we concentrate on this simple gauge for
developing an ML-MCTDH-based approach to purification.

\subsection{Overview of ML-MCTDH} \label{sec:ml-mctdh}

Here we give a brief overview of the ML-MCTDH ansatz and its equivalent
representation in terms of tree-tensor networks intended to provide a
self-contained and unified description within this work, while fixing some
nomenclature for the upcoming sections. It can be skipped by readers familiar
with ML-MCTDH. Note that here we use the position representation of the
wavefunction instead of the more general Dirac's notation, as this helps clarify
the different possible mode combinations, and is in line with the usual
nomenclature found in the MCTDH literature. 

The MCTDH ansatz expands a wavefunction of $F$ DOFs in a Hartree product basis
with expansion coefficients $A^1_{j_1,\dots,j_{p} }(t)$
\begin{align}
    \Psi(q_1,&\dots,q_F,t) = \\&\sum_{j_1}^{n_1}\cdots\sum_{j_p}^{n_p}  A^1_{j_1,\dots,j_{p} }(t)\cdot\varphi_{j_1}^{1;1}(Q_1^1,t)\cdot \dots \varphi_{j_p}^{1;p}(Q_{p}^1,t),\label{eq:mctdh-top}
\end{align}
where both the expansion coefficients and the Hartree products are
time-dependent. This ansatz corresponds to a Tucker format with core-tensor
$A^1_{j_1,\dots,j_{p}}(t)$. The functions
$\varphi_{j_\kappa}^{1;\kappa}(Q_\kappa^1,t)$ are called single-particle
functions (SPFs), and depend on a single coordinate $Q_\kappa=q_\kappa$ or a
subset of DOFs. The latter is referred to as a combined mode, e.g.,
$Q^1_1=\lbrace q_1,q_2\rbrace$, which turns $\varphi_{j_1}^{1;1}(Q_1^1,t)$ into
a two-dimensional function. This SPF can be either expanded in a
time-independent (product) basis, which is referred to as a primitive basis,
i.e., 
\begin{gather}
    \varphi_{m}^{1,1}(Q_1^1,t) = \sum_{i_{1}}^{N_{1}}\sum_{i_{2}}^{N_{2}} A_{m;i_{1},i_{2}}^{2;1}(t)\cdot  \chi^{2;1}_{i_1} (q_{1}) \chi^{2;2}_{i_2} (q_{2}).
\end{gather}
Alternatively, the SPF can be in turn expanded in a time-dependent SPF basis,
which is the basic principle of multilayer MCTDH (ML-MCTDH): Starting from the
expansion Eq.~\eqref{eq:mctdh-top}, the top-layer's many-dimensional SPFs are
expanded in lower-dimensional SPFs of the next, underlying layer. This
hierarchical expansion is again applied in the underlying layer, and repeated
until the expansion in the primitive, time-independent basis is reached. 

Thus, a SPF in the general ML case is given as
\begin{gather}
    \varphi^{z-1,\kappa_{l-1}}_m (Q^{z-1}_{\kappa_{l -1}}) = \sum_{j_1}^{n_1}\cdots\sum_{j_{p_{\kappa_l}}}^{n_{p_{\kappa_l}}}A^{z}_{m;j_1,\dots ,j_{p_{\kappa_l}}}(t)\prod_{\kappa_l} ^{p_{\kappa_l}} \varphi^{z,\kappa_{l}}_{j_{\kappa_l}}(Q^{z}_{\kappa_l}),
\end{gather}
where we have introduced the label $z=(l;\kappa_1,\dots,\kappa_{l-1})$ with
layer index $l$ to specify the position within the ML-expansion according to
Ref.~\onlinecite{vendrell2011multilayer}. The mode $Q^{z-1}_{\kappa_{l-1}}$ is the combined
mode of the following layer's modes, $Q^{z-1}_{\kappa_{l-1}}=\lbrace
Q^z_{1},\dots,Q^z_{p_{\kappa_l}} \rbrace$.
Note that orthonormality of SPFs
requires
\begin{equation}
    \label{eq:mctdh-ortho}
    \sum_{J}(A^{z}_{n;J})^*A^{z}_{m;J}=\delta_{nm},
\end{equation}
with combined index
$J=j_1,\dots ,j_{p_{\kappa_l}}$.

 While a direct expansion in a time-independent product basis would require
 $\prod_{\kappa=1}^F N_\kappa$ coefficients, the MCTDH and ML-MCTDH
 \emph{ansätze} yield a substantially more compact representation of the
 wavefunction. This is because, in general, the number of SPFs can be chosen
 lower than the number of underlying basis functions, owing to their flexibility
 as they are time-dependent. Hence, the propagation of a single, large
 coefficient tensor is broken down into propagating many lower-dimensional,
 manageable tensors.

The ML-MCTDH ansatz is visually represented as a tree graph
(cf.~Fig.~\ref{fig:1}) by representing the coefficient tensors
$A^{z}_{m;j_1,\dots ,j_{p_{\kappa_l}}}$ as nodes, and tensor indices as legs,
which connect two nodes when they are contracted over. Numbers next to legs
indicate the maximum value of the associated index, i.e., the number of SPFs or
primitive basis functions.
The ML-MCTDH ansatz then corresponds to a tree-tensor
network (TTN) with the particular orthonormality relation~\eqref{eq:mctdh-ortho}
obeyed by all nodes (for the topmost $l=m=1$, the usual normalization condition).
As an overview, equations of motion for this ansatz were derived and implemented by
Wang and Thoss~\cite{wang2003multilayer} and Manthe~\cite{manthe2008multilayer},
and further discussion and details about the commonly used
Heidelberg-package implementation were provided by Vendrell and
Meyer~\cite{vendrell2011multilayer}. The connection
between ML-MCTDH and TTNs has been explored in detail by Larsson~\cite{lar24:e2306881}
and alternative propagation schemes based on a more general normalization of the
intermediate tensors were discussed by Manthe~\cite{wei21:194108}.

\subsection{Compact tree representations of the ITS and PTS}\label{theo-its}

\begin{figure}
    \centering
    \includegraphics[width=\linewidth]{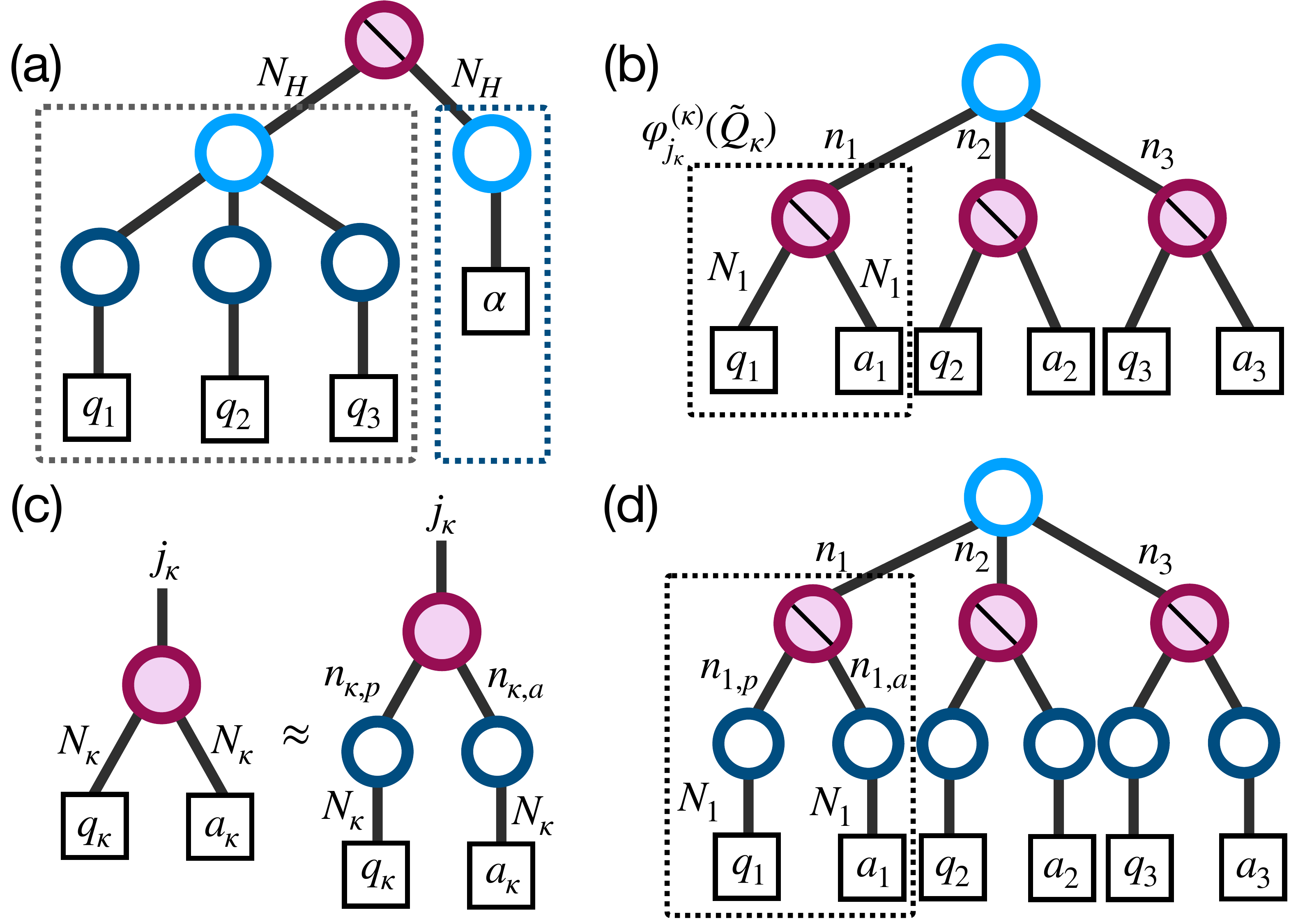}
    \caption{Possible ML-MCTDH trees for representing the ITS. (a) Combining full-dimensional physical and auxiliary spaces at the top-node. (b) Combining physical and auxiliary spaces referring to a single (possibly combined) mode at the bottom layer. (c) Mixed physical-auxiliary SPFs (STPFs) can be expanded in a time-dependent basis of separate physical and auxiliary sets of SPFs. With this expansion, the ITS is represented by ansatz~(d), where initially $n_{\kappa,p}=n_{\kappa,a}$. Diagonal lines denote diagonal coefficient tensors which are
    proportional to unit matrix as introduced around Eq.~\eqref{eq:one-ml-top}
    and Eq.~\eqref{eq:one-hartree-1}. In \emph{ansätze} (b) and (d), all nodes above the mixed physical-auxiliary layer [last layer in (b), second-to-last in (d)] are initially populated as rank-1 nodes.
    }
    \label{fig:1}
\end{figure}

Looking back at Eq.~\eqref{eq:one-pur-2} for the ITS, one may ask if there
exist efficient representations as an ML-MCTDH tree.
One may start, naively, by representing the physical and auxiliary systems as two
subtrees connected at the top node, indicated by dotted boxes in Fig.~\ref{fig:1}(a).
This corresponds to the ansatz 
\begin{gather}
    |\mathds{1}\rangle = \sum_{J,L}^{N_H} A_{JL}^1|\Phi_J\rangle|\tilde\alpha_L\rangle
     \label{eq:one-ml-top},
\end{gather}
where the top node $A_{JL}^1$ is equal to the $N_H$-dimensional unit
matrix, $A_{JL}^1=\delta_{JL}$.

In order for the $|\Phi_J\rangle$ basis states to cover the complete physical
Hilbert space, one has to set full ranks to all nodes in
any tensor decomposition
underlying it, which in practice results in the upper limit of the sum in
Eq.~\eqref{eq:one-ml-top} becoming the full dimension of the primitive space, $N_H$.
The single node representing the auxiliary subsystem can be chosen, for example
as a Euclidean set of orthonormal vectors $|\tilde\alpha_J\rangle$.
Clearly, this ansatz is very inefficient,
as it concentrates all entanglement between the physical and auxiliary system at
the top node.

Another possibility exists by breaking the auxiliary Hilbert space into a
product space of as many auxiliary subspaces as physical (or primitive)
degrees of freedom $F$,
\begin{align}
|\tilde\alpha_J\rangle &= |\tilde\alpha_{j_1\dots j_F}\rangle
                  =|\tilde\alpha_{j_1}^{(1)}\rangle \cdots |\tilde\alpha_{j_F}^{(F)}\rangle.
\end{align}
It turns out, the ITS can be exactly represented as
\begin{align}
    |\mathds{1}\rangle
    &= 
        \prod_{\kappa=1}^F
        \left(
            \sum_{j_\kappa}^{N_\kappa}
            |\chi_{j_{\kappa}}^{(\kappa)}\rangle 
            |\tilde\alpha_{j_{\kappa}}^{(\kappa)}\rangle
        \right)
        \label{eq:one-hartree-1}\\
    & =
        \sum_J^{N_H}|\Phi_J\rangle |\tilde\alpha_J\rangle.
        \label{eq:one-hartree-2} 
\end{align}
with $|\Phi_J\rangle
=\prod_{\kappa}|\chi_{j_{\kappa}}^{(\kappa)}\rangle$ and combined
index $J=(j_1,\dots,j_F)$.
Eq.~\eqref{eq:one-hartree-1} corresponds to a Hartree product of combined modes
containing one physical (primitive) degree of freedom and its auxiliary counterpart,
shown in Fig.~\ref{fig:1}(b), connected by diagonal nodes  (shown in purple).
Although relations \eqref{eq:one-hartree-1} and \eqref{eq:one-hartree-2} are strictly
equivalent, \eqref{eq:one-hartree-1} corresponds to the maximally compact
representation, exploiting the fact that the
overall entanglement between physical and auxiliary spaces in the
ITS can be factorized into the entanglement of each physical
degree of freedom and its auxiliary counterpart. This constitutes an
important result, thus enabling the practical application of
purification to general complex systems through compact representations
of the ITS and, subsequently, the PTS.

The ITS Hartree product can be trivially mapped onto any tree topology by
initially setting all node ranks above the bottom, diagonal layer to $1$. The
example tree topology in Fig.~\ref{fig:1}(b) corresponds to a two-layer MCTDH
tree.
Equation~\eqref{eq:map-dens-th} is then solved in practice by integrating
the imaginary-time Schrödinger equation
\begin{gather}
    \frac{\partial}{\partial \tau}|\psi(\tau)\rangle = -\hat H|\psi(\tau)\rangle
\end{gather}
up to $\tau=\beta/2$ with the (ML-)MCTDH algorithm\cite{wang_quantum-mechanical_2006,manthe_partition_2001,matzkies1998accurate}. 
Note that the partition function \eqref{eq:th-par} is given by the norm of the relaxed ITS,  $Z_\beta=\langle \mathds{1} e^{-\beta \hat{H}/2}|e^{-\beta \hat{H}/2}\mathds{1}\rangle$. By keeping the MCTDH wavefunction normalized during the imaginary-time propagation, the partition function in Eq.~\eqref{eq:map-dens-th} is hence automatically included in the PTS.

As we discuss in detail below, this imaginary-time propagation results in an
increase of the ranks of the intermediate and top tree nodes, which reflects the
existing correlation between physical degrees of freedom brought by the physical
Hamiltonian. The resulting PTS $|\psi_\beta\rangle$ at the target temperature is
then propagated using Eq.~\eqref{eq:tdse-pur} (and not Eq.~\eqref{eq:tdse-tf})
in real time to compute observables and correlation functions of interest.

\subsection{Thermal density of states from wavepacket propagation}

Lastly, it is interesting to note that the thermal density of states (TDOS)
\begin{align}
    p(E)=\frac{1}{Z_\beta}\mathrm{Tr}[e^{-\beta \hat H}\delta(E-\hat H)]
\end{align}
can be obtained straightforwardly in this formulation from the power spectrum
of the PTS, thus providing the Boltzmann distribution contained
in $|\psi_\beta\rangle$ from a regular spectral analysis.
This can be seen by
expressing $p(E)$ in the
time domain, 
\begin{align}
    p(E)&= \frac{1}{Z_\beta}\sum_n e^{-\beta E_n} \delta(E-E_n)\\
    &=\frac{1}{2\pi}\int_{-\infty}^\infty dt\frac{1}{Z_\beta}\sum_n e^{-\beta E_n} e^{i(E-E_n)t}\\
    &=\frac{1}{2\pi} \int_{-\infty}^\infty dt \,\,e^{iEt} \frac{1}{Z_\beta} \sum_n \langle n|e^{-\beta\hat H}e^{-i\hat Ht}|n\rangle\\
    &=\frac{1}{2\pi}\int_{-\infty}^\infty e^{iEt} \,\,
        \frac{1}{Z_\beta}\mathrm{Tr}\left[e^{-\frac{\beta}{2}\hat H}e^{-i\hat Ht}e^{-\frac{\beta}{2}\hat H}\right]dt\\
    &=\frac{1}{2\pi}\int_{-\infty}^\infty e^{iEt} a_\beta(t) dt,
\end{align}
thus rewriting the trace in the second-to-last equation using a purified density operator
in analogy to Eqs.~\eqref{eq:expr-pur-exp0} and \eqref{eq:expr-pur-expect},
\begin{gather}
   \frac{1}{Z_\beta}\mathrm{Tr}[e^{-\frac{\beta}{2}\hat H}e^{-i\hat Ht}e^{-\frac{\beta}{2}\hat H}]
   =\langle \psi_{\beta}(0)|\psi_{\beta}(t)\rangle=a_\beta(t),\label{eq:tdos-pur}
\end{gather}
where one sets $\hat A\to e^{-i\hat H t}$ %and $\hat B\to\hat 1$ 
in Eq.~\eqref{eq:expr-pur-exp0}.
As usual, noting that $a(-t)=\left[a(t)\right]^*$ for hermitian $\hat{H}$
yields $p(E)=\frac{1}{2\pi}\mathrm{Re}\int_{0}^\infty e^{iEt} a_\beta(t) dt$, and,
in practice, the finite propagation duration sets the Fourier-limited energy resolution of $p(E)$.
Examples of TDOS computed using this procedure are discussed in Section \ref{sec:results}.

\section{Implementation into ML-MCTDH}

\subsection{Pruning the finite-temperature tree}\label{sec:theo-prune}

At $\beta=0$, the full configurational space is populated with equal weights to
yield the ITS in Eq.~\eqref{eq:one-hartree-1}.  In the energy eigenbasis, many
states contribute negligibly to the canonical density operator at
finite-temperature due to exponential damping with $e^{-\beta E_n}$ with
eigenenergies $E_n$. Consequently, the number of required auxiliary
configurations in the PTS should substantially reduce compared to the exact ITS,
as the imaginary-time propagation to the target temperature proceeds.  This
motivates a dynamic pruning algorithm which removes unneeded physical and
auxiliary configurations during the ``cool-down'' of the ITS.

The procedure can be best explained by considering the ansatz in Fig.~\ref{fig:1}(d). 
There, the combined physical–auxiliary SPFs (``single thermal particle
functions'', STPFs) are expanded in separate sets of physical SPFs and auxiliary
SPFs [cf.~Fig.~\ref{fig:1}(c)], rather than being kept as a single combined mode
as in Fig.~\ref{fig:1}(b).
At $\beta=0$, all node ranks above the second-to-bottom diagonal layer
are set to 1, whereas $n_{\kappa,p} = n_{\kappa,a} = N_\kappa$, the size of
the $\kappa$-th physical space. The columns of the bottom-layer tensors (dark blue)
can be initialized as Euclidean vectors without loss of generality. 
Next, at each pre-defined imaginary-time interval $\Delta \beta$, the
bottom-layer tensors are pruned via transformation to natural orbitals, thus
ordering auxiliary and physical configurations by their natural populations.
Natural orbitals with natural populations larger than a pre-defined
threshold $\alpha_{\mathrm{thresh}}$ are kept and transformed back to their
original representation. After the pruning step the SPFs are
renormalized.
Note that the ranks of the nodes above the second-to-bottom layer grow during
the imaginary-time relaxation due to correlation in the Hamiltonian of the physical
system.

 When approaching the 0~K limit, $\beta\rightarrow \infty$, the on-the-fly
 pruning  automatically converges to the pure ground-state wavefunction, since
 the number of auxiliary configurations will be reduced to 1 and all higher-energy
 physical configurations decay exponentially.

\begin{figure}
   \centering
   \includegraphics[width=1.0\linewidth]{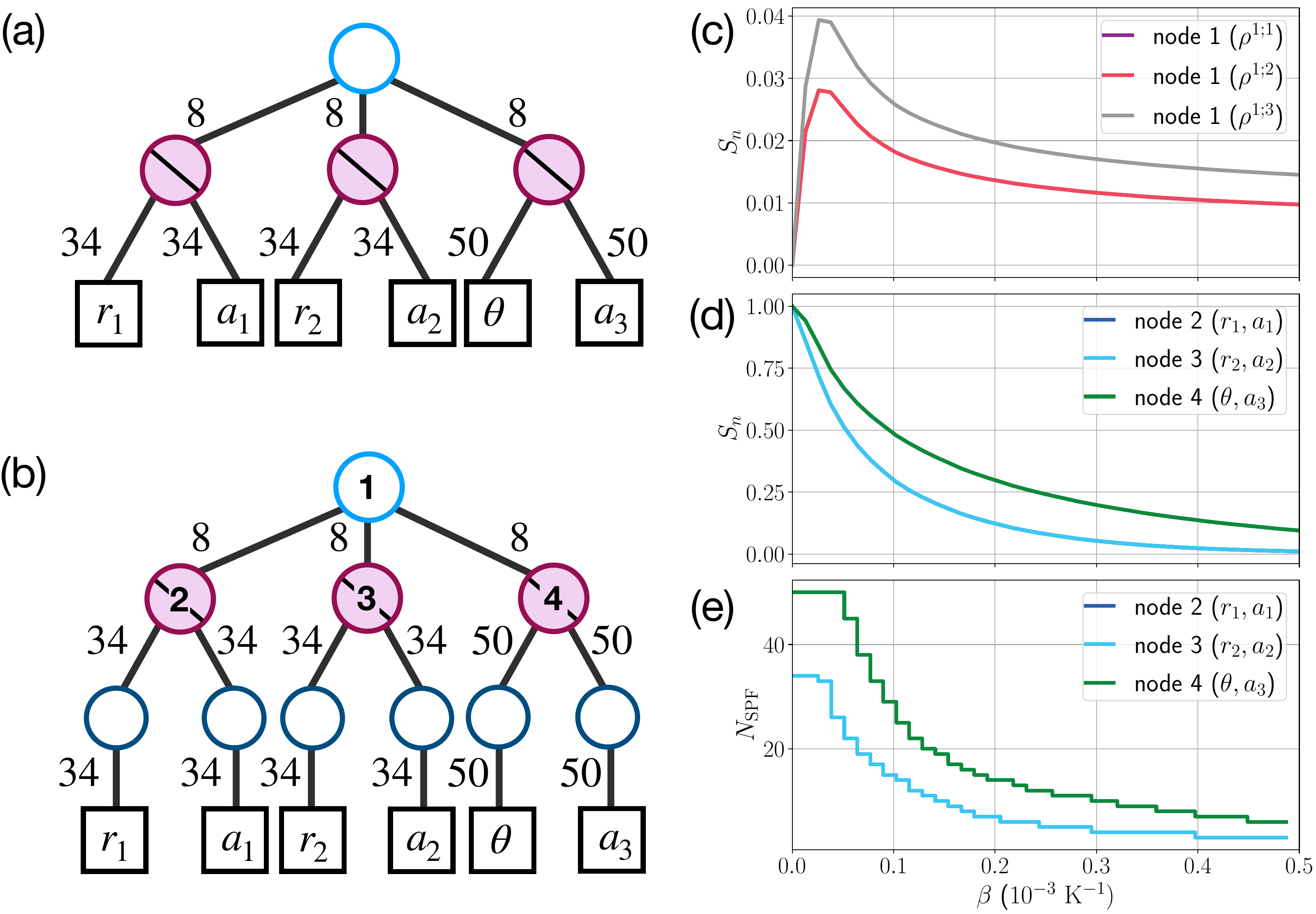}
   \caption{Thermal relaxation of the ITS for \ce{H2O}. (a) ML-MCTDH tree for
   ITS with combined physical-auxiliary modes at the bottom layer. (b) ML-MCTDH
   tree for ITS with separate SPFs for physical and auxiliary DOFs, compatible
   with pruning. (c)-(d) Normalized von Neumann entropies calculated at each leg
   of the top node [node 1 in (b)] and at the mixed physical-auxiliary nodes
   [node 2-4 in (b)]. (e) Dynamic pruning of SPFs, showing the number of
   physical and auxiliary SPFs at nodes 2-4 as a function of $\beta$. At each
   node, physical and auxiliary $N_{\mathrm{SPF}}(\beta)$ are on top of each
   other. A pruning algorithm with $\alpha_{\mathrm{thresh}}=10^{-5}$ and
   $\Delta\beta=0.01$~fs was used.}
   \label{fig:h2o-relax}
\end{figure}

Figure~\ref{fig:h2o-relax} tracks the damping of physical--auxiliary
configurations during the relaxation of \ce{H2O} by computing the normalized von
Neumann entropies at the STPF nodes [purple nodes in
Fig.~\ref{fig:h2o-relax}(b)]. With increasing $\beta$, these entropies decay
quickly, following an approximately exponential decay, and directly reflecting
the decreasing weights of the natural orbitals.  Consequently, the pruning
algorithm rapidly reduces the number of physical and auxiliary SPFs at the
thermal nodes [Fig.~\ref{fig:h2o-relax}(e)]. For large $\beta$, the pure state
limit $S_n\to 0$ is approached.
In contrast, computing $S_n$ at the top node reflects the build-up of
correlation among the STPFs caused by the correlated Hamiltonian of the physical
system. For large $\beta$, these entropies approach the value found for the
\ce{H2O} ground state, which exhibits some correlation between the stretching
and bending modes.
Both quantities, together with the achievable compression through pruning, are
analyzed in detail in Sec.~\ref{sec:h2o}.

At small $\beta$, the pruning algorithm cannot yet discard SPFs below the threshold $\alpha_\mathrm{thresh}$ due to insufficient damping [cf.~Fig.~\ref{fig:h2o-relax}(e)]. In practice, we hence find that the ansatz of Fig.~\ref{fig:1}(d), in which an extra layer separates the physical and auxiliary modes, is markedly less efficient at small $\beta$ than the ansatz of Fig.~\ref{fig:1}(b), where physical and auxiliary modes share a combined mode (see also Fig.~\ref{fig:1}(c) for how the two ansätze are related). Once pruning has removed a sufficient number of SPFs, however, the situation reverses: the ansatz in Fig.~\ref{fig:1}(d) becomes by far the more efficient representation, both during the remainder of the thermal relaxation and throughout the subsequent real-time propagation. We demonstrate this behavior for the 
\ce{H3O2-} system for the relaxation run (Supplementary Material, Fig.~S2) and real-time evolution in Sec.~\ref{sec:h3o2}.

This suggests an alternative pruning method in which an ITS with
primitive physical-auxiliary mode STPFs [Fig.~\ref{fig:1}(b)] is
cooled down initially to temperature $1/\beta$. This is in general faster
than a pruning run with the extra-layer ansatz, but one is left with bulkier
combined modes for the real-time propagation.
Therefore, an SVD transformation should be applied to the STPF nodes,
again only keeping the first $n_{\kappa,p}$ and $n_{\kappa,a}$
physical and auxiliary singular vectors, and then transfer these vectors to
an extra-layer ansatz for real-time propagation. This latter step has not yet
been implemented, but the expected
computational savings of this approach are estimated and discussed in
Sec.~\ref{sec:discussion}.

\section{Results}\label{sec:results}

All features needed for constructing and dynamically pruning the thermal ML-MCTDH
tree have been implemented in the Heidelberg MCTDH package and are
available in version 8.6.10\cite{mctdh:MLpackage}.
We apply our implementation
to the water molecule (\ce{H2O},
3D) and the hydrated hydroxide anion (\ce{H3O2-}, 9D). The low dimensionality of
the water molecule allows to benchmark the purification approach against
explicitly constructing the canonical density operator from precomputed energy
eigenstates. By propagating PTSs, we study the
thermal density of states (TDOS) and temperature effects on the IR
spectra of the fluxional \ce{H3O2-}
system.

\subsection{\ce{H2O}}\label{sec:h2o}
To test the accuracy of our approach, we compute the canonical density operator
of \ce{H2O}, using the electronic ground-state surface by Polyansky, Jensen and
Tennyson\cite{polyansky1994spectroscopically,polyansky1996potential}. For
reference, the first 105 energy eigenstates and energies (up to $\approx$
18000\,cm$^{-1}$) are obtained by block improved relaxation. This allows to
construct the canonical density operator in the energy eigenbasis using
Boltzmann weights $p_i=e^{-\beta E_i}$. 

\subsubsection{Thermal relaxation and pruning}

The PTS $|\psi_\beta\rangle$ is obtained by $\beta$-propagation of the ITS,
using either the two-layer tree in Fig.~\ref{fig:h2o-relax}(a) without pruning,
or the three-layer tree in Fig.~\ref{fig:h2o-relax}(b) with on-the-fly pruning.
In the latter case, we start with the exact ITS, i.e., as many SPFs as
underlying primitive basis functions are used. Further computational details can
be found in the Supplementary Material.

As expected, during the $\beta$-propagation of the ITS, the number of relevant
ensemble wavefunctions representing the canonical density operator decreases
quickly. We quantify this decay by computing the normalized von Neumann entropy
of reduced density matrices, $S_n$. Note that the entropy is normalized by its
maximum value $S_n=S/\mathrm{ln}\,n_{\kappa_p}$, where
$n_{\kappa,p}=n_{\kappa,a}$ is the number of physical basis functions. Thus, a
maximally mixed state corresponds to $S_n=1$, a pure state to $S_n=0$.

During the ``cool down'' of the ITS, $S_n$ decreases quickly
[cf.~Fig.~\ref{fig:h2o-relax}(d)], indicating the shrinking rank of the physical
and auxiliary reduced density matrix. As temperature decreases, the bending mode
$\theta$ maintains a more mixed reduced density matrix compared to the bond
stretches. Due to its lower excitation energy, the bending modes
remain thermally populated at temperatures where the stretching modes
have already relaxed to an almost pure ground state. 

Accordingly, dynamic pruning monotonically reduces the number of physical and
auxiliary SPFs with increasing $\beta$, following closely the decay of the
entropy. At $\beta=0.5\times 10^{-3}~\mathrm{K}^{-1}$, it is sufficient to retain about 10\% of the initial SPFs at
a threshold of $\alpha_\mathrm{thresh}=10^{-5}$. Fig.~\ref{fig:h2o-relax}(e)
also illustrates that in the limit of low temperatures, the pruning algorithm
converges to a pure state ML-MCTDH wavefunction by reducing the auxiliary
subspace to a single configuration. 

While entropies of the reduced physical and auxiliary density matrices decay
with $\beta$, entropies calculated at the top-node, i.e., in the space of the
STPFs, show a different behavior. They reflect the inter-mode correlation
averaged over all ensemble wavefunctions. At $\beta=0$, the ITS is fully
uncorrelated ($S_n=0$), since the $|\Phi_J\rangle$'s in Eq.~\eqref{eq:one-hartree-2} are
Hartree products. In the $\beta\to\infty$ limit the entropy decays to that of
the correlated vibrational ground state of water. Due to physical correlation,
the top-node entropies approach a
non-zero value for large $\beta$ in Fig.~\ref{fig:h2o-relax}(c). In the
intermediate $\beta$ regime, a manifold of highly excited, correlated states is
thermally populated, resulting in a maximum of the entropies.

We conclude that the dynamic pruning generally allows for a more compact
representation of the STPFs, exploiting the exponential damping of higher-energy
configurations during thermal relaxation. During real-time evolution, the
auxiliary subsystem remains stationary, making further extension of the pruned
SPF bases unnecessary.
However, the ranks of physical density matrices and upper-layer nodes are not
conserved during real-time propagation. Consequently, the SPF spaces must be
chosen sufficiently large such that inter-mode correlations in all wavefunctions
of the ensemble are accurately captured. This requires typically more physical
SPFs than a converged 0~K tree for the same system.

\subsubsection{Thermal observables}
In order to assess the accuracy of the purification and pruning method, we start
by comparing thermal expectation values at 500~K, 1000~K and 2000~K in
Table~\ref{tab:tab1}. Absolute values are given for the eigenstate reference,
while the two purification-based methods are characterized by their deviation
from it. Since the observables differ strongly in absolute magnitude, relative
errors quoted below are given with respect to the reference value and, for the
stretching and bending observables, additionally with respect to the thermal
variation $\Delta X_{\mathrm{therm}}=X(2000\,\mathrm{K})-X(500\,\mathrm{K})$.

Overall, both purification variants reproduce the Boltzmann-weighted averages
over eigenstates very well. The mean thermal energy $\langle \hat H\rangle_{\beta}$,
given with respect to the zero-point energy, deviates by at most $0.92$\,cm$^{-1}$
(pur-full) and $0.73$\,cm$^{-1}$ (pur-pruned) at 2000~K, corresponding to relative
errors below $0.1\,\%$. Deviations of the stretching and bending expectation
values do not exceed $6\times10^{-4}$ in the respective units, i.e., relative
errors below $0.02\,\%$ for all temperatures and modes, or below $2\,\%$ (for $\langle \theta\rangle_\beta$)
of the thermal variation of the corresponding observable between 500~K and
2000~K. The deviations of the two purification variants are of the same order of
magnitude throughout.

All deviations grow with temperature, as expected: at higher temperatures the
thermal ensemble contains more highly excited states, which are more sensitive to the level of convergence of the underlying representation.
We stress that the eigenstate reference is itself not exact, since it is obtained from
block improved relaxation and converged to a finite number of states.
The pruning error is controlled by $\alpha_{\mathrm{thresh}}$ and can be reduced
by lowering the threshold. In the Supplementary Material (Fig.~S1), the pruning
error and the corresponding number of coefficients (i.e., the achievable
compression of the PTS) are examined  as a function of
$\alpha_{\mathrm{thresh}}$. The threshold chosen in Tab.~\ref{tab:tab1},
$\alpha_{\mathrm{thresh}}=10^{-5}$, is already sufficiently low such that full
and pruned PTS perform equally well within the overall numerical accuracy of
the simulations.

\begin{table}
\caption{Thermal observables for \ce{H2O} from eigenstates obtained by improved
relaxation (``eigen''), purification with a full-dimensional combined
physical--auxiliary grid [ansatz in Fig.~\ref{fig:1}(b), ``pur-full''], and
dynamically pruned PTS based on the ansatz in Fig.~\ref{fig:1}(d)
(``pur-pruned''), with $\alpha_{\mathrm{thresh}}=10^{-5}$. For the two
purification-based methods, the deviation from the eigenstate reference is given
in the unit of the respective observable.
$\langle\hat H\rangle_{\beta}$ is measured from the zero-point energy; $ r_2$
is equivalent to $ r_1$ by symmetry and is not listed.}
\begin{ruledtabular}
\begin{tabular}{lccc}
 & 500\,K & 1000\,K & 2000\,K \\ \hline
\multicolumn{4}{l}{$\langle \hat H\rangle_{\beta}$\,(cm$^{-1}$)}\\
\quad eigen      & 16.59      & 217.19     & 1335.03    \\
\quad pur-full   & $+0.01$    & $+0.11$    & $+0.92$    \\
\quad pur-pruned & $-0.01$    & $+0.06$    & $+0.73$    \\[3pt]
\multicolumn{4}{l}{$\langle \hat r_{1}\rangle_{\beta}$\,(bohr)}\\
\quad eigen      & 1.84337    & 1.84413    & 1.85036    \\
\quad pur-full   & $+0.00000$ & $+0.00002$ & $+0.00007$ \\
\quad pur-pruned & $+0.00000$ & $+0.00001$ & $+0.00007$ \\[3pt]
\multicolumn{4}{l}{$\langle \hat\theta\rangle_{\beta}$\,(rad)}\\
\quad eigen      & 1.82298    & 1.82511    & 1.83239    \\
\quad pur-full   & $+0.00003$ & $+0.00008$ & $+0.00014$ \\
\quad pur-pruned & $-0.00001$ & $+0.00017$ & $+0.00013$ \\[3pt]
\multicolumn{4}{l}{$\langle \hat r_{1}^{2}\rangle_{\beta}$\,(bohr$^{2}$)}\\
\quad eigen      & 3.41499    & 3.41804    & 3.44401    \\
\quad pur-full   & $-0.00001$ & $+0.00004$ & $+0.00027$ \\
\quad pur-pruned & $-0.00001$ & $+0.00004$ & $+0.00026$ \\[3pt]
\multicolumn{4}{l}{$\langle \hat\theta^{2}\rangle_{\beta}$\,(rad$^{2}$)}\\
\quad eigen      & 3.34782    & 3.36073    & 3.40674    \\
\quad pur-full   & $+0.00010$ & $+0.00030$ & $+0.00053$ \\
\quad pur-pruned & $+0.00009$ & $+0.00028$ & $+0.00050$ \\
\end{tabular}
\end{ruledtabular}
\label{tab:tab1}
\end{table}

\begin{figure}
    \centering
    \includegraphics[width=0.9\linewidth]{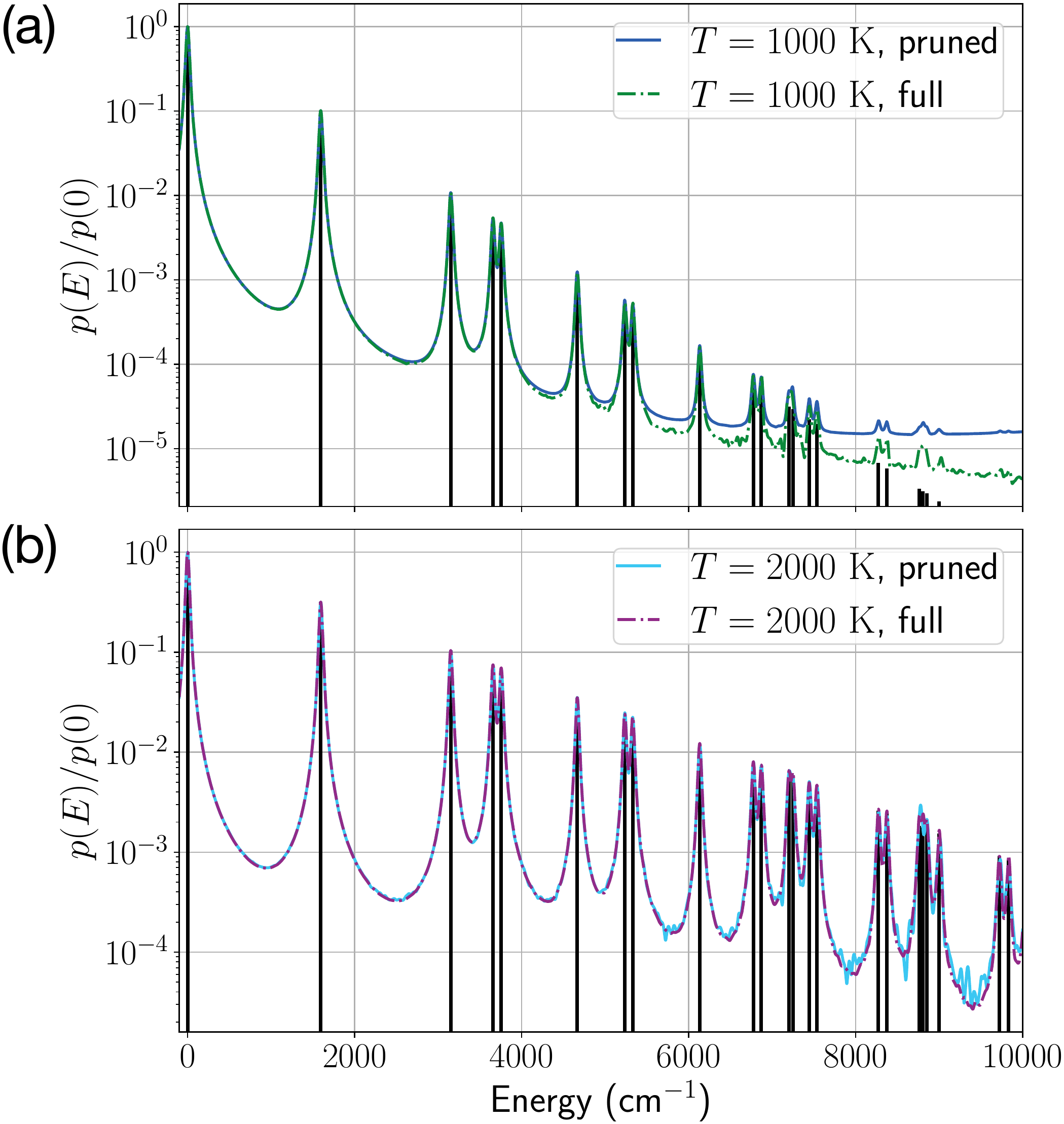}
    \caption{TDOS for \ce{H2O} at (a) 1000~K and (b) 2000~K, computed from the PTS via Eq.~\eqref{eq:tdos-pur}. The PTS is either obtained from the ITS in Fig.~\ref{fig:1}(b) (``full'') or via pruning starting from Fig.~\ref{fig:1}(d) (``pruned''). For the pruning, a threshold $\alpha_\mathrm{thresh}=10^{-5}$ is used. Note that the TDOS was normalized such that the ground state peak height is set to 1. This way, $p(E)$ reflects the bare Boltzmann weights, $\sum_n\delta(E-E_n)e^{-\beta E_n}$ (without the partition function) at finite energy resolution. For reference, the stick TDOS is given, with eigenenergies $E_n$ computed from improved relaxation.}
    \label{fig:fig3}
\end{figure}

The TDOS in Fig.~\ref{fig:fig3} was obtained by propagating the PTS in real time for 800~fs. Even for the low-dimensional \ce{H2O} system, the timings of this propagation indicate favorable performance of the pruned PTS: Instead of evolving a STPF on a physical-auxiliary primitive grid with 1156 ($r_1$, $r_2$) and 2500 ($\theta$) grid points, an expansion in a physical-auxiliary SPF space with 56 ($r_1$, $r_2$) and 294 ($\theta$) (at 2000~K) functions captures the TDOS equally well. At 1000~K, the size of the physical-auxiliary space can be pruned even further to 20 ($r_1$, $r_2$) and 56 ($\theta$), corresponding to approximately 2~\% of the original physical-auxiliary primitive grid sizes.  

The compressed dimensionality at \emph{each} mixed physical-auxiliary node decreases
the total size of the ML-MCTDH wavefunction substantially: compared to combined
physical-auxiliary primitive grids [ansatz in Fig.~\ref{fig:h2o-relax}(a)], the
total number of coefficients is reduced by 93~\% at 1000~K (2762 vs. 41572
coefficients), and 82~\% at 2000~K (19618 vs. 109392). 

Consequently, real-time evolution of the pruned tree is significantly faster.
The CPU wall time decreases from 12~h to 49~min at 1000~K, and from about 142~h to
17.5~h at 2000~K. Comparisons between pruned and ``full'' PTS time-evolution
were performed with an equal number of SPFs in the top-layer, and a variable
mean-field integration scheme with a Runge-Kutta (RK8) integrator.

Both TDOSs, obtained with and without pruning, agree well with the
stick-spectrum obtained from the eigenstates. This indicates that the pruning
parameter $\alpha_{\mathrm{thresh}}$ was chosen sufficiently low, such that only
configurations with negligible weight were removed, not affecting the relevant
thermally-populated part of the molecular DOS. Eigenenergies are reproduced from
the time-evolved purified density operator whereas intensities deviate from the
Boltzmann distribution above approximately 7000~cm$^{-1}$ at 1000~K. The
influence of these states on the thermal observables in Tab.~\ref{tab:tab1},
which average over this distribution, is marginal due to their low weight
($\sim10^{-5}$). 
Also note that the low-intensity contributions to $p(E)$ are sensitive to
numerical noise, and are represented at lower accuracy compared to the dominant
configurations. As soon as these states gain more weight at higher temperature
[cf.~Fig.~\ref{fig:fig3}(b)], much better agreement with the eigenstate results
is achieved.

\subsection{\ce{H3O2-}} \label{sec:h3o2}
We now turn to the more complex, nine-dimensional \ce{H3O2-} system. By calculating the temperature-dependent IR-spectrum of this prototypical, highly flexible molecule, we demonstrate the applicability of the ML-MCTDH purification approach to larger, strongly anharmonic systems with multiple low-frequency modes. To this end, the potential energy and dipole surfaces from Refs.~\onlinecite{pelaez2014full,pelaez2017infrared} were refitted using the Monte-Carlo Canonical Polyadic Decomposition (MCCPD) implementation in the Heidelberg MCTDH package.

\subsubsection{Assessing the performance of dynamic pruning}

We start by computing the purified canonical density operators of \ce{H3O2-} for
various temperatures between 50~K and 250~K with dynamical pruning. The pruning
threshold is set to $\alpha_{\mathrm{thresh}}=10^{-6}$. This results in a
drastic reduction of the auxiliary spaces at the target temperature compared to
the exact ITS. For instance, at 250~K auxiliary SPF numbers are reduced to about
1-10\% of the full primitive grid size for all modes except the torsion angle
$\phi$ ($\approx 70\%$). Similar to the bending angle of \ce{H2O}, this rotational
low-energy excitation results in many thermally populated states compared to the
higher-energy vibrations. At 50~K, auxiliary spaces are compressed even further
to $\approx$ 1\% of their original size, except the torsion which is pruned to 29\% of the original SPF
number. More details on the ML-MCTDH trees before and after pruning can be found
in the Supporting Information (Figs.~S3 and S4).

Compared to the relaxation with doubled primitive grids  [cf. ansatz in
Fig.~\ref{fig:1}(b)], the thermal relaxation of the ITS with doubled SPF spaces
[cf. ansatz in Fig.~\ref{fig:1}(c)-(d)] is much slower. CPU timings are shown in
Tab.~\ref{tab:2}. The increased computational effort can be traced back almost
exclusively to the first few initial integration steps (cf.~Fig.~S2 in the
Supplementary Material), where the full auxiliary-physical SPF space is
uniformly populated. During the first time steps, pruning is not yet possible
due to insufficient damping of the high-energy configurations.
However, as soon as high-energy
configurations can be removed during the imaginary-time evolution, this approach
becomes faster compared to combined primitive physical and auxiliary grids.

The pronounced increase in CPU time for the relaxation is redeemed by a
significantly reduced effort in the real-time propagation: the much more compact
ML-MCTDH wavefunction is reflected in an approximately ten times shorter CPU
time in Tab.~\ref{tab:2}.
These computational savings in real-time propagation for a given
temperature become significant as one needs to
run the thermal relaxation only once to obtain $|\psi_\beta\rangle$, but usually
several propagation runs are required to extract thermal observables. For
instance, calculating the IR spectrum of \ce{H3O2-} requires propagation of
$|\psi_{\beta}(t)\rangle$, $\hat\mu_x|\psi_{\beta}(t)\rangle$,
$\hat\mu_y|\psi_{\beta}(t)\rangle$ and $\hat\mu_z|\psi_{\beta}(t)\rangle$,
corresponding to the thermal density operator and the x-,y- and z-polarized
components of the dipole-operated density operator.

\subsubsection{Temperature-dependent IR spectrum}

\begin{figure*}
    \centering
    \includegraphics[width=1.0\linewidth]{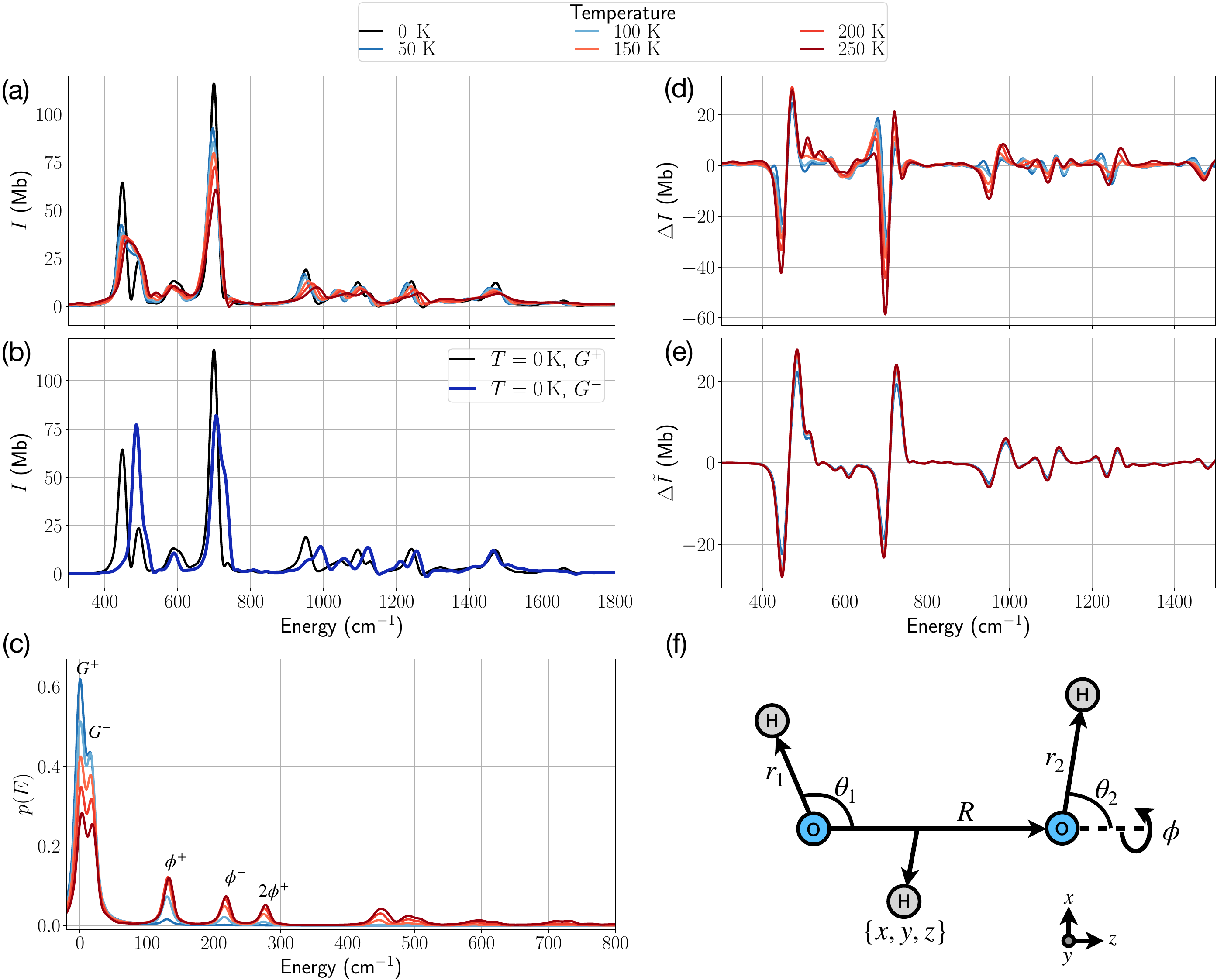}
    \caption{Thermal spectra of the 9D hydrated hydroxide anion \ce{H3O2-}. (a) IR absorption spectra for various temperatures from 0~K to 250~K. (b) IR absorption spectra at 0~K, starting from the pure even-parity ($G^+$) or odd-parity ($G^-$) ground state. (c) TDOS obtained from the real-time evolved PTS, assignments based on Ref.~\onlinecite{pelaez2014full}. The TDOS is normalized and scaled such that the peak heights correspond to the probabilities $p_n=e^{-\beta E_n}/Z_{\beta}$. (d) Temperature-dependent difference spectra $\Delta I = I_\beta-I_{G^{+}}$ with respect to the 0~K $G^+$ IR absorption spectrum. (e) Approximate difference spectrum $\Delta\tilde{I}$ based on the two-state model introduced in the main text. (f) \ce{H3O2-} valence coordinates. }
    \label{fig:h3o2-spec}
\end{figure*}

The temperature-dependent IR absorption spectrum in Fig.~\ref{fig:h3o2-spec}(a)
is obtained as the average over the $x$, $y$, $z$-polarized linear absorption
spectra, $I_\beta(E)=\frac{1}{3}(I_{\beta,x}(E)+I_{\beta,y}(E)+I_{\beta,z}(E))$,
which are computed from the thermal dipole-dipole correlation functions
$\langle\mu_i(t)\mu_i(0)\rangle_\beta$ from Eq.~\eqref{eq:th-mu-mu} for
$i=x,y,z$. Correlation functions were computed up to
$t_{\mathrm{final}}=1600$\,fs.

Three pronounced temperature-dependent features can be identified: broadening
and merging of the double peak at 400-500\,cm$^{-1}$, suppression of the
high-intensity band around 700\,cm$^{-1}$, and blue-shift of higher-energy bands
above 900\,cm$^{-1}$. The difference spectra $\Delta I$ between the
finite-temperature and 0~K spectra in Fig.~\ref{fig:h3o2-spec}(d) reveals rather
intricate temperature-induced changes in these regions.

Interpretation of these spectral signatures is not straightforward.  At $T=0$~K,
all IR-active transitions originate from the absolute ground vibrational state,
rendering the spectrum a direct probe of excited-state energies relative to a
reference ground-state energy.
However, at finite temperature, IR absorption occurs from a manifold of
thermally populated initial states; the IR spectrum contains all dipole
allowed transitions (energy differences) between thermally populated initial
states and corresponding final states.
This makes the interpretation of spectral features of highly flexible and
anharmonic systems at finite temperature particularly challenging.

We examine this situation more closely in the \ce{H3O2-} system by computing the
TDOS between 50 and 250~K. While at 50~K only two states above
the ground-state have Boltzmann weights greater than 1~\%, several vibrationally
excited states up to 700\,cm$^{-1}$ are substantially populated at higher
temperatures [Fig.~\ref{fig:h3o2-spec}(c)]. At 250~K, 5 distinct peaks in the
TDOS carry weights of about $5$--$30\%$ and together account for
roughly $80\%$ of the total Boltzmann weight; they are therefore expected to
dominate the thermal IR absorption spectrum. In order of increasing energy, they
are assigned to the even-parity ground state ($G^+$), odd-parity ground state
($G^-$), even fundamental of the torsion ($\phi^+$), odd fundamental of the
torsion ($\phi^-$), as well as the even-parity overtone of the torsion ($2\phi^+$)
\cite{pelaez2014full,pelaez2017infrared}. See Fig.~\ref{fig:h3o2-spec}(f) for the
definition of the torsion angle, and the remaining valence coordinates.

Even and odd parity labels arise from the symmetry along the torsion angle,
where odd states vanish at $\phi=\pi$ and even ones do not. Importantly, this
results in a small tunneling splitting of the ground state,
$\omega_{\mathrm{tunnel}}=17$\,cm$^{-1}$. Owing to this rather small energy
difference, the $G^+$ and $G^-$ states are almost equally populated, and their
combined weight dominates the thermally populated manifold. The ratio of their
populations shows little temperature-dependence, saturating quickly above 100~K.
It should be stressed, however, that this dominance is far from exclusive:
already at 250~K the $G^\pm$ doublet carries only about one half of the total
Boltzmann weight, with the majority of the remainder distributed over the
low-lying torsional ladder ($\phi^\pm$, $2\phi^+$), which lies within $k_bT$ of
the ground state. Any reduced description based on the tunneling doublet alone
is therefore expected to be qualitative, and we will return to the fingerprints
of the neglected states below.

Since $G^+$ and $G^-$ are the dominant contributions to the initially populated
manifold, we expect that the thermal spectra in Fig.~\ref{fig:h3o2-spec} can be
-- at least qualitatively -- understood as a weighted sum of the pure-state
spectra of the even and odd parity ground states and their corresponding
Boltzmann factor. These spectra are shown in Fig.~\ref{fig:h3o2-spec}(b),
differing clearly in the three thermally-active regions identified above. The merging bands between 400-500~cm$^{-1}$ and blue-shifting band of the bridging hydrogen's (BH) $z$-motion are a result of emerging hot bands close to the fundamental transition of the OH rocking and BH z-motion.

This can be rationalized by considering the selection rule for the $z$-polarized
dipole transitions, which dominate over the $x$ and $y$-polarized components of
$I(E)$. Based on the molecular symmetry group of the floppy \ce{H3O2-}, $\mu_z$
induces only transitions between states with equal $\pm$ symmetry labels so that
only $G^+\!\to\!+$ and $G^-\!\to\!-$ transitions carry
intensity\cite{vendrell2007full,bunker2018fundamentals}. Each fundamental therefore appears once
per pure-state spectrum in Fig.~\ref{fig:h3o2-spec}(b), exclusively exciting one
of the two components of the torsional doublets.

For a given band, a hot band appears in the $G^-$ spectrum displaced by
$\Delta-\omega_{\mathrm{tunnel}}$, where $\Delta$ denotes the tunneling
splitting of the excited state.
Due to the anharmonicity, the energy difference between the $\pm$ doublets
depends on the number of vibrational quanta in the remaining modes, reflecting
how far a given excitation promotes or hinders the torsional tunneling. 

The double peak at 400-500\,cm$^{-1}$ consists of the OH rocking
($\varphi_1+\varphi_2$) and OO stretching fundamentals roughly at 450 and
490\,cm$^{-1}$ in $I_{G^+}$\cite{pelaez2014full}. Exciting the rocking mode
roughly doubles the $\pm$-splitting ($\Delta\approx 40$\,cm$^{-1}$),
blue-shifting it to 470\,cm$^{-1}$ in $I_{G^-}$, whereas the OO stretch is
essentially a spectator ($\Delta\approx\omega_{\mathrm{tunnel}}$) and remains at
490\,cm$^{-1}$. The odd-parity rocking band thus falls within the range of the
even-parity peaks, which leads to an apparent merging of the peaks in
Fig.~\ref{fig:h3o2-spec}(a) when the $G^-$ state is populated with increasing
temperature.

The high-intensity band at $\approx 700$\,cm$^{-1}$ is assigned to the BH
$z$-motion, the even counterpart lies $\approx 35$~cm$^{-1}$ above the
odd-symmetry state\cite{pelaez2014full}, resulting in a blue-shifted hot band in
$I_{G^-}$ which has lower intensity compared to $I_{G^+}$.
The weighted
average of both spectra results in a slight blue shifting of the BH $z$ band
center. In the more dense region above 900~cm$^{-1}$ similar trends can be
observed, their exact assignment lies however beyond the scope of this work.

To quantify the spectral impact of vibrationally excited states beyond the
two-state model, we compare the temperature-dependent difference spectra $\Delta
I = I_{\beta} - I_{G^{+}}$ with the approximated $\Delta \tilde{I} = p_+
I_{G^{+}} + p_- I_{G^{-}} - I_{G^{+}} =
\frac{\mathrm{e}^{-\beta\omega_{\mathrm{tunnel}}}}
{1+\mathrm{e}^{-\beta\omega_{\mathrm{tunnel}}}}(I_{G^-}-I_{G^+})$ based on the
two-state model including only the ground state doublet. In
Fig.~\ref{fig:h3o2-spec}(d)-(e), the overall shape is well described by the
reduced model, with larger deviations at the $z$-motion band of the bridging
hydrogen -- unsurprisingly, given that this is precisely the band whose
intensity is redistributed over several eigenstates in $I_{G^-}$. Notably, the
approximated $\Delta\tilde{I}$ displays much weaker dependence on the
temperature. The quick saturation of the thermal population ratio between
tunneling-split states explains this behavior: since $\omega_{\mathrm{tunnel}}
\ll k_bT$ over the entire range considered, $p_-/p_+ \to 1$ in the two-state model, and
$\Delta\tilde{I}$ becomes an almost temperature-independent difference of two fixed
spectra.
In contrast, the stronger temperature dependence observed in absorption
spectra from the purified thermal density operator in
Fig.~\ref{fig:h3o2-spec}(d) suggests the involvement of higher-lying vibrational
states, whose populations are more sensitive to temperature.

Lastly, quantitatively reproducing the spectrum by summing pure-state spectra
would require computing and propagating all eigenstates up to
$\approx1000$\,cm$^{-1}$ [cf.~TDOS in Fig.~\ref{fig:h3o2-spec}(c)].
This is still feasible for the nine-dimensional \ce{H3O2-}, but the
DOS for low-energy excitations increases rapidly in larger systems, rendering the
sum-over-states approach computationally prohibitive. 
This growth is captured quantitatively by direct state-counting schemes such as
the Beyer--Swinehart algorithm and its Stein--Rabinovitch extension
\cite{beyer1973algorithm,stein1973accurate}, which show a near-exponential rise
of $p(E)$ with both energy and dimensionality.

Simulating IR absorption at higher temperatures hence becomes increasingly
difficult, as the Boltzmann distribution extends into regions of significantly
higher vibrational state density. This makes the purification approach
favorable, since it treats the thermal ensemble in a single propagation and its
cost is set by the entanglement of the purified state rather than by $p(E)$
itself.

\begin{table}
\caption{Performance of PTS ansatz with combined primitive physical-auxiliary
modes (``primitive''), and a pruned PTS with a non-primitive physical-auxiliary
node (``non-prim.'') for \ce{H3O2-}. CPU-timings (total time) for ITS relaxation
and propagation of $\hat{\mu}_x|\psi_\beta\rangle$. The ML-trees used for this
comparison are shown in Fig.~S4 of the Supplementary Material.}
\begin{ruledtabular}
\begin{tabular}{cccc}
 & &  \multicolumn{2}{c}{timing\,[h]}\\
 calculation & $T$\,(K) & primitive ($t^\mathrm{CPU}_\mathrm{prim}$) & non-prim. ($t^\mathrm{CPU}_\mathrm{pruned}$) \\\hline
 relaxation & 150 & 70 &   662      \\
  relaxation & 50 & 144 &   667     \\
 propagation & 150  &  5777 \footnote{extrapolated from a 400\,fs propagation}  & 644\\
 propagation & 50 & 2644  & 234 \\ 
\end{tabular}
\end{ruledtabular}
\label{tab:2}
\end{table}

\section{Discussion}\label{sec:discussion}

Let us recapitulate the main features of our implementation of purification
within ML-MCTDH: 
First, the PTS $|\psi_\beta\rangle$ is obtained from an \emph{exact}, compact
representation of the ITS via imaginary time propagation. At $\beta=0$, the
ML-MCTDH wavefunction is initialized as a Hartree product of
infinite-temperature STPFs, while at finite-temperature the multilayer
tree-tensor network efficiently represents correlations among STPFs. 

Second, the algorithm prunes the STPFs space of corresponding
physical and auxiliary modes. Removing
configurations with negligible natural weights from the expansion of STPFs,
substantially reduces the overall number of coefficients which need to be
propagated in real time. 

Third, during real-time propagation the auxiliary modes stay constant in time
(unit auxiliary Hamiltonian) as the system evolves under the physical
Hamiltonian, which does not act on the auxiliary modes. 

Finally, both imaginary-time and real-time evolution are computed with the
standard ML-MCTDH equations of motion for relaxation and propagation,
respectively.  Observables (expectation values, correlation functions,\dots) are
evaluated in exact analogy to expressions for pure states.
We note in passing, that efficient application of the physical Hamiltonian to
the purified ML-tree is automatically guaranteed in the Heidelberg MCTDH
implementation, as unit operators on the auxiliary modes are flagged in advance,
and are not explicitly multiplied\cite{vendrell2011multilayer}. 

Overall, these features result in several advantages over existing
finite-temperature methods within the MCTDH framework.  The sampling approach
requires careful checking of convergence with respect to the number of samples.
Depending on the system and temperature, large samples may be required.
Moreover, convergence of a specific observable does not guarantee convergence of
all thermal observables.

To illustrate this, we have conducted a random-phase sampling of the thermal
ground-state of \ce{H3O2-} at 250~K (cf.~Fig.~S5 in the Supplementary Material).
The one-dimensional density along the OO stretch converges quickly, with a batch
size of 100 samples already providing a good description. However, the density
along the torsion angle is significantly harder to capture accurately with the
sampling approach. Even $\sim$~1000 samples do not reproduce exactly the
symmetric distribution around $\phi=\pi$, lower batch sizes break this
symmetry noticeably. This makes reliable simulations of the
temperature-dependent IR spectrum challenging, since it is 
dominated by the initial thermal distribution of torsional states.

We have found the pruning scheme decisive for tractable real-time evolution of
PTSs, because purification alone doubles the number of degrees of freedom and
thereby squares the entire primitive space. Although (ML-)MCTDH is far less
sensitive than grid-based methods to the size of the primitive grids, the
squaring of all primitive grids still inflates the wavefunction significantly.
When physical primitive modes are themselves already combined, which is
typically the case for large and highly flexible systems~\cite{schroder2022coupling}, this becomes even more problematic.  At the same time, mode
combination of physical modes has proven to be a valuable strategy in MCTDH and
ML-MCTDH for arriving at compact, near-optimal tree
structures\cite{mendive2023optimal}. Pruning reaches a favorable compromise: it
allows one to retain such mode combinations, which reflect strongly correlated
groups of physical degrees of freedom, while associating them with an auxiliary
space that is much smaller than the primitive space of the combined mode. 

From our analysis of thermal correlations in Fig.~\ref{fig:h2o-relax}, we expect
the pruning to be advantageous in the regime of low to moderate temperatures. At
high temperatures, nodes at which physical and auxiliary SPFs meet will retain a
high rank. In this case, directly combining the primitive physical and auxiliary
grids avoids the additional overhead of an extra layer while propagating SPF
spaces similar to the primitive grid size. At cryogenic temperatures, the
canonical density operator is dominated by a few low-lying eigenstates. In this
regime, a sampling or eigenstate-based approach is expected to be more
favorable. 

We stress that a large portion of thermal quantum systems of interest operate
precisely in this intermediate regime, where the thermal‑node rank is modest yet
non‑trivial and pruning yields its largest savings. A prominent example is
excitation‑energy transfer in photosynthetic light‑harvesting complexes, whose
relevant vibrational modes lie in the tens‑to‑hundreds of cm$^{-1}$ range,
comparable to the thermal energy at physiological temperatures ($k_bT\approx
200$~cm$^{-1}$ at 300~K)\cite{schulze2015explicit,ishizaki2009theoretical}.

A few words comparing the purification approach with pruning and
(ML-)$\rho$-MCTDH are in order here. Both approaches effectively double the
number of DOFs, but differ conceptually in what is propagated, and how.
The ansatz with combined physical and auxiliary primitive modes [cf.~Fig.~\ref{fig:1}(b)] is closely
related to the (ML-)$\rho$-MCTDH type-I ansatz: identifying the auxiliary
primitive index of mode $\kappa$ with the bra-coordinate $Q_\kappa'$, a single
thermal particle function $\varphi^{(\kappa)}(Q_\kappa,\tilde{Q}_\kappa)$ plays
the same role as a hermitian single-particle density operator
$\sigma^{(\kappa)}(Q_\kappa,Q_\kappa')$.  In $\rho$-MCTDH the doubled ansatz
maps one-to-one onto a density operator evolved under the LvN equation, whereas
our purification is an ordinary (ML-)MCTDH wave-function propagation in which
only the physical subsystem is propagated [Eq.~\eqref{eq:tdse-pur}].
Consequently, hermiticity and positivity of
$\rho=\mathrm{Tr}_{\mathrm{aux}}|\psi_\beta\rangle\langle\psi_\beta|$ hold by
construction as $\rho$ is the partial trace of a pure-state projector, and trace
conservation follows from the norm conservation of the underlying (pruned)
ML-MCTDH propagation. In contrast, (ML-)$\rho$-MCTDH must enforce hermiticity by
imposing a hermitian initial state.

Moreover, since the purification ML-MCTDH does not map one-to-one on the
time-evolved density operator but only after taking the partial trace, removing
unnecessary configurations at low temperature becomes easily possible.  This is
in stark contrast to $\rho$-MCTDH, where the bra index is itself propagated
under the LvN equation and carries dynamical information, so its low-weight
configurations in the bottom tensors retain their full dimension $N_\kappa
\times N_\kappa$ irrespective of the target temperature and cannot be discarded
without losing physical content.
Together with the more involved Liouville-space equations of
motion~\cite{van_haeften_propagating_2023}, this leads to substantially larger
coefficient tensors. 

As Tab.~\ref{tab:2} shows, the resulting pruned thermal states clearly
outperform the unpruned combined-mode ansatz of Fig.~\ref{fig:1}(b) throughout
the real-time propagation. Since the wavefunction ansatz in Fig.~\ref{fig:1}(b)
is as costly as a $\rho$-MCTDH type-I ansatz (which usually performs better than
type-II\cite{raab_multiconfiguration_1999,van_haeften_propagating_2023}),
comparable savings relative to $\rho$-MCTDH and its multilayer generalization can be anticipated.

Within the current implementation, there are two bottlenecks: the initial
relaxation of the ITS in the region of small $\beta$ is rather slow due to the
overhead of representing the infinite-temperature STPFs in the format of
Fig.~\ref{fig:1}(c), as discussed in Sec.~\ref{sec:theo-prune}. Furthermore,
compared to a pure state (0~K) calculation, increased numbers of SPFs may be
required for the real-time propagations, with the exception of the auxiliary
SPFs. Regarding the latter, determining the appropriate number of SPFs and the
optimal ML-tree structure which minimizes this number is a central requirement
for any ML-MCTDH calculation. Recent research focused on automatizing these
steps\cite{larsson_ttns_2019,mendive2023optimal,mendive2017towards}, and could
be extended to finite-temperature tree-tensor networks. 

The first bottleneck could be addressed by replacing the dynamic pruning method
with the faster thermal relaxation, at least for small $\beta$, followed by a
direct transformation to a compact, pruned tree with an extra layer at the
bottom of the form in Fig.~\ref{fig:1}(d).
Assuming negligible overhead of this one-time transformation (essentially
diagonalizing the reduced density matrix of the combined physical-auxiliary pair
and keeping natural orbitals above a threshold), we estimate the savings of this
approach from the timings in Tab.~\ref{tab:2}. For 150~K we obtain for ratio of
total relaxation plus propagation CPU times
$(t^\mathrm{CPU}_\mathrm{prim,rlx}+t^\mathrm{CPU}_\mathrm{pruned,prop})/(t^\mathrm{CPU}_\mathrm{pruned,rlx}+t^\mathrm{CPU}_\mathrm{pruned,prop})=0.55$,
i.e., a speed-up by about a factor of 2 compared to the current dynamic pruning
approach. Alternatively, the ITS could be constructed in a suitable truncated
basis, such as eigenfunctions of the uncorrelated part of $\hat{H}$.

\section{Conclusion}

We have implemented a new scheme for treating finite-temperature systems with
ML-MCTDH based on purification. This approach relies on augmenting the physical
DOFs with an auxiliary subsystem, which allows the mapping of a density operator
to a single wavefunction in the augmented space. The doubled number of DOFs is
still handled efficiently by using the multilayer generalization of MCTDH.
Additionally, we have implemented a dynamic pruning scheme during imaginary-time
propagation to achieve a more compact representation of the purified thermal state 
within the ML-MCTDH framework. 

Most importantly, this purification approach enables the computation of a wide range of
thermal observables by propagating a \emph{single} ML-MCTDH wavefunction. We
have examined here, as proof-of-concept, linear absorption spectra, thermal
densities of states (TDOS), and static expectation values. By transferring
pure-state expressions to the purified thermal state, the methodology can be
easily extended to study more complex, experimentally relevant systems and properties,
including pump-probe or non-linear spectra at finite temperatures exploiting the efficiency of ML-MCTDH.
A further advantage is the direct applicability to both model Hamiltonians, and
complex, \emph{ab initio}-derived potential energy surfaces.

Simulations of the highly flexible 9D \ce{H3O2-} cluster highlight the
capability of the purification-pruning approach to handle complex
high-dimensional systems on \emph{ab initio} Hamiltonians.
The results point to an additional layer of complexity in the interpretation of
thermal spectra of fluxional systems: While 0~K spectra provide a direct probe
of excited states relative to the absolute ground state, finite temperature
involves transitions from a manifold of thermally populated initial states. This
leads to partially overlapping peaks with a different initial state origin,
temperature-dependent band shifts, and broadening. In large systems, the
situation becomes even more involved due to the increased number of
low-frequency modes. This shall motivate the future development of methods for analysis
and assignment of thermal spectra. Our purification scheme provides an important
step forward in the simulation of such spectral signatures, while circumventing
completely the explicit computation of eigenstates in spectrally dense regions.

 In conclusion, this work lays the foundation for studying high-dimensional,
 general, and complex quantum systems at finite-temperature based on a single ML-MCTDH
 wavefunction, by combining the simplicity and flexibility of purification with
 the compactness and efficiency of a tree-tensor network method. Ultimately, we
 anticipate that these results will expand the scope of finite-temperature
 quantum simulations and drive further methodological developments, allowing for
 deeper insights into the quantum dynamics of thermalized molecules, clusters
 and materials.

\section*{Supplementary Material}

The Supplementary Material contains ML-tree structures and primitive grids for \ce{H2O} and \ce{H3O2-} calculations, additional results on the pruning error and CPU timings, and a comparison between the random-phase sampling and purification approach for \ce{H3O2-} at 250~K.

\begin{acknowledgments}
    The authors are grateful to Prof. Hans-Dieter Meyer for his assistance
    in implementing the approach within the Heidelberg MCTDH package.
    The authors acknowledge support by the state of Baden-Württemberg through bwHPC
    and the German Research Foundation (DFG) through grant INST 35/1597-1 FUGG. 
\end{acknowledgments}

\bibliography{purification}

%aipnum4-2.bst 2019-01-14 (MD) hand-edited version of apsrev4-1.bst
\begin{thebibliography}{65}%
\makeatletter
\providecommand \@ifxundefined [1]{%
 \@ifx{#1\undefined}
}%
\providecommand \@ifnum [1]{%
 \ifnum #1\expandafter \@firstoftwo
 \else \expandafter \@secondoftwo
 \fi
}%
\providecommand \@ifx [1]{%
 \ifx #1\expandafter \@firstoftwo
 \else \expandafter \@secondoftwo
 \fi
}%
\providecommand \natexlab [1]{#1}%
\providecommand \enquote  [1]{``#1''}%
\providecommand \bibnamefont  [1]{#1}%
\providecommand \bibfnamefont [1]{#1}%
\providecommand \citenamefont [1]{#1}%
\providecommand \href@noop [0]{\@secondoftwo}%
\providecommand \href [0]{\begingroup \@sanitize@url \@href}%
\providecommand \@href[1]{\@@startlink{#1}\@@href}%
\providecommand \@@href[1]{\endgroup#1\@@endlink}%
\providecommand \@sanitize@url [0]{\catcode `\\12\catcode `\$12\catcode
  `\&12\catcode `\#12\catcode `\^12\catcode `\_12\catcode `\%12\relax}%
\providecommand \@@startlink[1]{}%
\providecommand \@@endlink[0]{}%
\providecommand \url  [0]{\begingroup\@sanitize@url \@url }%
\providecommand \@url [1]{\endgroup\@href {#1}{\urlprefix }}%
\providecommand \urlprefix  [0]{URL }%
\providecommand \Eprint [0]{\href }%
\providecommand \doibase [0]{https://doi.org/}%
\providecommand \selectlanguage [0]{\@gobble}%
\providecommand \bibinfo  [0]{\@secondoftwo}%
\providecommand \bibfield  [0]{\@secondoftwo}%
\providecommand \translation [1]{[#1]}%
\providecommand \BibitemOpen [0]{}%
\providecommand \bibitemStop [0]{}%
\providecommand \bibitemNoStop [0]{.\EOS\space}%
\providecommand \EOS [0]{\spacefactor3000\relax}%
\providecommand \BibitemShut  [1]{\csname bibitem#1\endcsname}%
\let\auto@bib@innerbib\@empty
%</preamble>
\bibitem [{\citenamefont {Tuckerman}\ \emph {et~al.}(1997)\citenamefont
  {Tuckerman}, \citenamefont {Marx}, \citenamefont {Klein},\ and\ \citenamefont
  {Parrinello}}]{tuckerman1997quantum}%
  \BibitemOpen
  \bibfield  {author} {\bibinfo {author} {\bibfnamefont {M.~E.}\ \bibnamefont
  {Tuckerman}}, \bibinfo {author} {\bibfnamefont {D.}~\bibnamefont {Marx}},
  \bibinfo {author} {\bibfnamefont {M.~L.}\ \bibnamefont {Klein}},\ and\
  \bibinfo {author} {\bibfnamefont {M.}~\bibnamefont {Parrinello}},\ }\bibfield
   {title} {\enquote {\bibinfo {title} {On the quantum nature of the shared
  proton in hydrogen bonds},}\ }\href@noop {} {\bibfield  {journal} {\bibinfo
  {journal} {Science}\ }\textbf {\bibinfo {volume} {275}},\ \bibinfo {pages}
  {817--820} (\bibinfo {year} {1997})}\BibitemShut {NoStop}%
\bibitem [{\citenamefont {Hern{\'a}ndez de~la Pe{\~n}a}\ and\ \citenamefont
  {Kusalik}(2005)}]{hernandez2005temperature}%
  \BibitemOpen
  \bibfield  {author} {\bibinfo {author} {\bibfnamefont {L.}~\bibnamefont
  {Hern{\'a}ndez de~la Pe{\~n}a}}\ and\ \bibinfo {author} {\bibfnamefont
  {P.~G.}\ \bibnamefont {Kusalik}},\ }\bibfield  {title} {\enquote {\bibinfo
  {title} {Temperature dependence of quantum effects in liquid water},}\
  }\href@noop {} {\bibfield  {journal} {\bibinfo  {journal} {Journal of the
  American Chemical Society}\ }\textbf {\bibinfo {volume} {127}},\ \bibinfo
  {pages} {5246--5251} (\bibinfo {year} {2005})}\BibitemShut {NoStop}%
\bibitem [{\citenamefont {Romero}\ \emph {et~al.}(2014)\citenamefont {Romero},
  \citenamefont {Augulis}, \citenamefont {Novoderezhkin}, \citenamefont
  {Ferretti}, \citenamefont {Thieme}, \citenamefont {Zigmantas},\ and\
  \citenamefont {Van~Grondelle}}]{romero2014quantum}%
  \BibitemOpen
  \bibfield  {author} {\bibinfo {author} {\bibfnamefont {E.}~\bibnamefont
  {Romero}}, \bibinfo {author} {\bibfnamefont {R.}~\bibnamefont {Augulis}},
  \bibinfo {author} {\bibfnamefont {V.~I.}\ \bibnamefont {Novoderezhkin}},
  \bibinfo {author} {\bibfnamefont {M.}~\bibnamefont {Ferretti}}, \bibinfo
  {author} {\bibfnamefont {J.}~\bibnamefont {Thieme}}, \bibinfo {author}
  {\bibfnamefont {D.}~\bibnamefont {Zigmantas}},\ and\ \bibinfo {author}
  {\bibfnamefont {R.}~\bibnamefont {Van~Grondelle}},\ }\bibfield  {title}
  {\enquote {\bibinfo {title} {Quantum coherence in photosynthesis for
  efficient solar-energy conversion},}\ }\href@noop {} {\bibfield  {journal}
  {\bibinfo  {journal} {Nature physics}\ }\textbf {\bibinfo {volume} {10}},\
  \bibinfo {pages} {676--682} (\bibinfo {year} {2014})}\BibitemShut {NoStop}%
\bibitem [{\citenamefont {Scholes}\ \emph {et~al.}(2017)\citenamefont
  {Scholes}, \citenamefont {Fleming}, \citenamefont {Chen}, \citenamefont
  {Aspuru-Guzik}, \citenamefont {Buchleitner}, \citenamefont {Coker},
  \citenamefont {Engel}, \citenamefont {Van~Grondelle}, \citenamefont
  {Ishizaki}, \citenamefont {Jonas} \emph {et~al.}}]{scholes2017using}%
  \BibitemOpen
  \bibfield  {author} {\bibinfo {author} {\bibfnamefont {G.~D.}\ \bibnamefont
  {Scholes}}, \bibinfo {author} {\bibfnamefont {G.~R.}\ \bibnamefont
  {Fleming}}, \bibinfo {author} {\bibfnamefont {L.~X.}\ \bibnamefont {Chen}},
  \bibinfo {author} {\bibfnamefont {A.}~\bibnamefont {Aspuru-Guzik}}, \bibinfo
  {author} {\bibfnamefont {A.}~\bibnamefont {Buchleitner}}, \bibinfo {author}
  {\bibfnamefont {D.~F.}\ \bibnamefont {Coker}}, \bibinfo {author}
  {\bibfnamefont {G.~S.}\ \bibnamefont {Engel}}, \bibinfo {author}
  {\bibfnamefont {R.}~\bibnamefont {Van~Grondelle}}, \bibinfo {author}
  {\bibfnamefont {A.}~\bibnamefont {Ishizaki}}, \bibinfo {author}
  {\bibfnamefont {D.~M.}\ \bibnamefont {Jonas}}, \emph {et~al.},\ }\bibfield
  {title} {\enquote {\bibinfo {title} {Using coherence to enhance function in
  chemical and biophysical systems},}\ }\href@noop {} {\bibfield  {journal}
  {\bibinfo  {journal} {Nature}\ }\textbf {\bibinfo {volume} {543}},\ \bibinfo
  {pages} {647--656} (\bibinfo {year} {2017})}\BibitemShut {NoStop}%
\bibitem [{\citenamefont {Perlik}\ \emph {et~al.}(2014)\citenamefont {Perlik},
  \citenamefont {Lincoln}, \citenamefont {Sanda},\ and\ \citenamefont
  {Hauer}}]{perlik2014distinguishing}%
  \BibitemOpen
  \bibfield  {author} {\bibinfo {author} {\bibfnamefont {V.}~\bibnamefont
  {Perlik}}, \bibinfo {author} {\bibfnamefont {C.}~\bibnamefont {Lincoln}},
  \bibinfo {author} {\bibfnamefont {F.}~\bibnamefont {Sanda}},\ and\ \bibinfo
  {author} {\bibfnamefont {J.}~\bibnamefont {Hauer}},\ }\bibfield  {title}
  {\enquote {\bibinfo {title} {Distinguishing electronic and vibronic coherence
  in 2d spectra by their temperature dependence},}\ }\href@noop {} {\bibfield
  {journal} {\bibinfo  {journal} {The Journal of Physical Chemistry Letters}\
  }\textbf {\bibinfo {volume} {5}},\ \bibinfo {pages} {404--407} (\bibinfo
  {year} {2014})}\BibitemShut {NoStop}%
\bibitem [{\citenamefont {Orenstein}\ and\ \citenamefont
  {Millis}(2000)}]{orenstein2000advances}%
  \BibitemOpen
  \bibfield  {author} {\bibinfo {author} {\bibfnamefont {J.}~\bibnamefont
  {Orenstein}}\ and\ \bibinfo {author} {\bibfnamefont {A.}~\bibnamefont
  {Millis}},\ }\bibfield  {title} {\enquote {\bibinfo {title} {Advances in the
  physics of high-temperature superconductivity},}\ }\href@noop {} {\bibfield
  {journal} {\bibinfo  {journal} {Science}\ }\textbf {\bibinfo {volume}
  {288}},\ \bibinfo {pages} {468--474} (\bibinfo {year} {2000})}\BibitemShut
  {NoStop}%
\bibitem [{\citenamefont {Simon}(1997)}]{simon1997superconductivity}%
  \BibitemOpen
  \bibfield  {author} {\bibinfo {author} {\bibfnamefont {A.}~\bibnamefont
  {Simon}},\ }\bibfield  {title} {\enquote {\bibinfo {title} {Superconductivity
  and chemistry},}\ }\href@noop {} {\bibfield  {journal} {\bibinfo  {journal}
  {Angewandte Chemie International Edition in English}\ }\textbf {\bibinfo
  {volume} {36}},\ \bibinfo {pages} {1788--1806} (\bibinfo {year}
  {1997})}\BibitemShut {NoStop}%
\bibitem [{\citenamefont {Duan}\ \emph {et~al.}(2017)\citenamefont {Duan},
  \citenamefont {Prokhorenko}, \citenamefont {Cogdell}, \citenamefont {Ashraf},
  \citenamefont {Stevens}, \citenamefont {Thorwart},\ and\ \citenamefont
  {Miller}}]{duan2017nature}%
  \BibitemOpen
  \bibfield  {author} {\bibinfo {author} {\bibfnamefont {H.-G.}\ \bibnamefont
  {Duan}}, \bibinfo {author} {\bibfnamefont {V.~I.}\ \bibnamefont
  {Prokhorenko}}, \bibinfo {author} {\bibfnamefont {R.~J.}\ \bibnamefont
  {Cogdell}}, \bibinfo {author} {\bibfnamefont {K.}~\bibnamefont {Ashraf}},
  \bibinfo {author} {\bibfnamefont {A.~L.}\ \bibnamefont {Stevens}}, \bibinfo
  {author} {\bibfnamefont {M.}~\bibnamefont {Thorwart}},\ and\ \bibinfo
  {author} {\bibfnamefont {R.~D.}\ \bibnamefont {Miller}},\ }\bibfield  {title}
  {\enquote {\bibinfo {title} {Nature does not rely on long-lived electronic
  quantum coherence for photosynthetic energy transfer},}\ }\href@noop {}
  {\bibfield  {journal} {\bibinfo  {journal} {Proceedings of the National
  Academy of Sciences}\ }\textbf {\bibinfo {volume} {114}},\ \bibinfo {pages}
  {8493--8498} (\bibinfo {year} {2017})}\BibitemShut {NoStop}%
\bibitem [{\citenamefont {Engel}\ \emph {et~al.}(2007)\citenamefont {Engel},
  \citenamefont {Calhoun}, \citenamefont {Read}, \citenamefont {Ahn},
  \citenamefont {Man{\v{c}}al}, \citenamefont {Cheng}, \citenamefont
  {Blankenship},\ and\ \citenamefont {Fleming}}]{engel2007evidence}%
  \BibitemOpen
  \bibfield  {author} {\bibinfo {author} {\bibfnamefont {G.~S.}\ \bibnamefont
  {Engel}}, \bibinfo {author} {\bibfnamefont {T.~R.}\ \bibnamefont {Calhoun}},
  \bibinfo {author} {\bibfnamefont {E.~L.}\ \bibnamefont {Read}}, \bibinfo
  {author} {\bibfnamefont {T.-K.}\ \bibnamefont {Ahn}}, \bibinfo {author}
  {\bibfnamefont {T.}~\bibnamefont {Man{\v{c}}al}}, \bibinfo {author}
  {\bibfnamefont {Y.-C.}\ \bibnamefont {Cheng}}, \bibinfo {author}
  {\bibfnamefont {R.~E.}\ \bibnamefont {Blankenship}},\ and\ \bibinfo {author}
  {\bibfnamefont {G.~R.}\ \bibnamefont {Fleming}},\ }\bibfield  {title}
  {\enquote {\bibinfo {title} {Evidence for wavelike energy transfer through
  quantum coherence in photosynthetic systems},}\ }\href@noop {} {\bibfield
  {journal} {\bibinfo  {journal} {Nature}\ }\textbf {\bibinfo {volume} {446}},\
  \bibinfo {pages} {782--786} (\bibinfo {year} {2007})}\BibitemShut {NoStop}%
\bibitem [{\citenamefont {Cao}\ \emph {et~al.}(2020)\citenamefont {Cao},
  \citenamefont {Cogdell}, \citenamefont {Coker}, \citenamefont {Duan},
  \citenamefont {Hauer}, \citenamefont {Kleinekath{\"o}fer}, \citenamefont
  {Jansen}, \citenamefont {Man{\v{c}}al}, \citenamefont {Miller}, \citenamefont
  {Ogilvie} \emph {et~al.}}]{cao2020quantum}%
  \BibitemOpen
  \bibfield  {author} {\bibinfo {author} {\bibfnamefont {J.}~\bibnamefont
  {Cao}}, \bibinfo {author} {\bibfnamefont {R.~J.}\ \bibnamefont {Cogdell}},
  \bibinfo {author} {\bibfnamefont {D.~F.}\ \bibnamefont {Coker}}, \bibinfo
  {author} {\bibfnamefont {H.-G.}\ \bibnamefont {Duan}}, \bibinfo {author}
  {\bibfnamefont {J.}~\bibnamefont {Hauer}}, \bibinfo {author} {\bibfnamefont
  {U.}~\bibnamefont {Kleinekath{\"o}fer}}, \bibinfo {author} {\bibfnamefont
  {T.~L.}\ \bibnamefont {Jansen}}, \bibinfo {author} {\bibfnamefont
  {T.}~\bibnamefont {Man{\v{c}}al}}, \bibinfo {author} {\bibfnamefont {R.~D.}\
  \bibnamefont {Miller}}, \bibinfo {author} {\bibfnamefont {J.~P.}\
  \bibnamefont {Ogilvie}}, \emph {et~al.},\ }\bibfield  {title} {\enquote
  {\bibinfo {title} {Quantum biology revisited},}\ }\href@noop {} {\bibfield
  {journal} {\bibinfo  {journal} {Science Advances}\ }\textbf {\bibinfo
  {volume} {6}},\ \bibinfo {pages} {eaaz4888} (\bibinfo {year}
  {2020})}\BibitemShut {NoStop}%
\bibitem [{\citenamefont {Scholes}\ \emph {et~al.}(2011)\citenamefont
  {Scholes}, \citenamefont {Fleming}, \citenamefont {Olaya-Castro},\ and\
  \citenamefont {Van~Grondelle}}]{scholes2011lessons}%
  \BibitemOpen
  \bibfield  {author} {\bibinfo {author} {\bibfnamefont {G.~D.}\ \bibnamefont
  {Scholes}}, \bibinfo {author} {\bibfnamefont {G.~R.}\ \bibnamefont
  {Fleming}}, \bibinfo {author} {\bibfnamefont {A.}~\bibnamefont
  {Olaya-Castro}},\ and\ \bibinfo {author} {\bibfnamefont {R.}~\bibnamefont
  {Van~Grondelle}},\ }\bibfield  {title} {\enquote {\bibinfo {title} {Lessons
  from nature about solar light harvesting},}\ }\href@noop {} {\bibfield
  {journal} {\bibinfo  {journal} {Nature chemistry}\ }\textbf {\bibinfo
  {volume} {3}},\ \bibinfo {pages} {763--774} (\bibinfo {year}
  {2011})}\BibitemShut {NoStop}%
\bibitem [{\citenamefont {May}\ and\ \citenamefont
  {K{\"u}hn}(2023)}]{may2023charge}%
  \BibitemOpen
  \bibfield  {author} {\bibinfo {author} {\bibfnamefont {V.}~\bibnamefont
  {May}}\ and\ \bibinfo {author} {\bibfnamefont {O.}~\bibnamefont {K{\"u}hn}},\
  }\href@noop {} {\emph {\bibinfo {title} {Charge and energy transfer dynamics
  in molecular systems}}}\ (\bibinfo  {publisher} {John Wiley \& Sons},\
  \bibinfo {year} {2023})\BibitemShut {NoStop}%
\bibitem [{\citenamefont {Manthe}(2008)}]{manthe2008multilayer}%
  \BibitemOpen
  \bibfield  {author} {\bibinfo {author} {\bibfnamefont {U.}~\bibnamefont
  {Manthe}},\ }\bibfield  {title} {\enquote {\bibinfo {title} {A multilayer
  multiconfigurational time-dependent hartree approach for quantum dynamics on
  general potential energy surfaces},}\ }\href@noop {} {\bibfield  {journal}
  {\bibinfo  {journal} {The Journal of chemical physics}\ }\textbf {\bibinfo
  {volume} {128}} (\bibinfo {year} {2008})}\BibitemShut {NoStop}%
\bibitem [{\citenamefont {Wang}\ and\ \citenamefont
  {Thoss}(2003)}]{wang2003multilayer}%
  \BibitemOpen
  \bibfield  {author} {\bibinfo {author} {\bibfnamefont {H.}~\bibnamefont
  {Wang}}\ and\ \bibinfo {author} {\bibfnamefont {M.}~\bibnamefont {Thoss}},\
  }\bibfield  {title} {\enquote {\bibinfo {title} {Multilayer formulation of
  the multiconfiguration time-dependent hartree theory},}\ }\href@noop {}
  {\bibfield  {journal} {\bibinfo  {journal} {The Journal of chemical physics}\
  }\textbf {\bibinfo {volume} {119}},\ \bibinfo {pages} {1289--1299} (\bibinfo
  {year} {2003})}\BibitemShut {NoStop}%
\bibitem [{\citenamefont {Vendrell}\ and\ \citenamefont
  {Meyer}(2011)}]{vendrell2011multilayer}%
  \BibitemOpen
  \bibfield  {author} {\bibinfo {author} {\bibfnamefont {O.}~\bibnamefont
  {Vendrell}}\ and\ \bibinfo {author} {\bibfnamefont {H.-D.}\ \bibnamefont
  {Meyer}},\ }\bibfield  {title} {\enquote {\bibinfo {title} {Multilayer
  multiconfiguration time-dependent hartree method: Implementation and
  applications to a henon--heiles hamiltonian and to pyrazine},}\ }\href@noop
  {} {\bibfield  {journal} {\bibinfo  {journal} {The Journal of Chemical
  Physics}\ }\textbf {\bibinfo {volume} {134}} (\bibinfo {year}
  {2011})}\BibitemShut {NoStop}%
\bibitem [{\citenamefont {Vendrell}, \citenamefont {Gatti},\ and\ \citenamefont
  {Meyer}(2009)}]{vendrell2009strong}%
  \BibitemOpen
  \bibfield  {author} {\bibinfo {author} {\bibfnamefont {O.}~\bibnamefont
  {Vendrell}}, \bibinfo {author} {\bibfnamefont {F.}~\bibnamefont {Gatti}},\
  and\ \bibinfo {author} {\bibfnamefont {H.-D.}\ \bibnamefont {Meyer}},\
  }\bibfield  {title} {\enquote {\bibinfo {title} {Strong isotope effects in
  the infrared spectrum of the zundel cation},}\ }\href@noop {} {\bibfield
  {journal} {\bibinfo  {journal} {Angewandte Chemie International Edition}\
  }\textbf {\bibinfo {volume} {48}},\ \bibinfo {pages} {352--355} (\bibinfo
  {year} {2009})}\BibitemShut {NoStop}%
\bibitem [{\citenamefont {Schr{\"o}der}\ \emph {et~al.}(2022)\citenamefont
  {Schr{\"o}der}, \citenamefont {Gatti}, \citenamefont {Lauvergnat},
  \citenamefont {Meyer},\ and\ \citenamefont
  {Vendrell}}]{schroder2022coupling}%
  \BibitemOpen
  \bibfield  {author} {\bibinfo {author} {\bibfnamefont {M.}~\bibnamefont
  {Schr{\"o}der}}, \bibinfo {author} {\bibfnamefont {F.}~\bibnamefont {Gatti}},
  \bibinfo {author} {\bibfnamefont {D.}~\bibnamefont {Lauvergnat}}, \bibinfo
  {author} {\bibfnamefont {H.-D.}\ \bibnamefont {Meyer}},\ and\ \bibinfo
  {author} {\bibfnamefont {O.}~\bibnamefont {Vendrell}},\ }\bibfield  {title}
  {\enquote {\bibinfo {title} {The coupling of the hydrated proton to its first
  solvation shell},}\ }\href@noop {} {\bibfield  {journal} {\bibinfo  {journal}
  {Nature Communications}\ }\textbf {\bibinfo {volume} {13}},\ \bibinfo {pages}
  {6170} (\bibinfo {year} {2022})}\BibitemShut {NoStop}%
\bibitem [{\citenamefont {{Mendive-Tapia}}\ \emph {et~al.}(2026)\citenamefont
  {{Mendive-Tapia}}, \citenamefont {Schran}, \citenamefont {Das}, \citenamefont
  {Gatti}, \citenamefont {Schr{\"o}der}, \citenamefont {Marx},\ and\
  \citenamefont {Vendrell}}]{men26:1}%
  \BibitemOpen
  \bibfield  {author} {\bibinfo {author} {\bibfnamefont {D.}~\bibnamefont
  {{Mendive-Tapia}}}, \bibinfo {author} {\bibfnamefont {C.}~\bibnamefont
  {Schran}}, \bibinfo {author} {\bibfnamefont {B.}~\bibnamefont {Das}},
  \bibinfo {author} {\bibfnamefont {F.}~\bibnamefont {Gatti}}, \bibinfo
  {author} {\bibfnamefont {M.}~\bibnamefont {Schr{\"o}der}}, \bibinfo {author}
  {\bibfnamefont {D.}~\bibnamefont {Marx}},\ and\ \bibinfo {author}
  {\bibfnamefont {O.}~\bibnamefont {Vendrell}},\ }\bibfield  {title} {\enquote
  {\bibinfo {title} {Deciphering the infrared spectrum of the hydrated proton
  using full-dimensional quantum dynamics},}\ }\href
  {https://doi.org/10.1038/s41557-026-02209-3} {\bibfield  {journal} {\bibinfo
  {journal} {Nature Chemistry}\ ,\ \bibinfo {pages} {1--8}} (\bibinfo {year}
  {2026})}\BibitemShut {NoStop}%
\bibitem [{\citenamefont {Wang}\ \emph {et~al.}(2011)\citenamefont {Wang},
  \citenamefont {Pshenichnyuk}, \citenamefont {H{\"a}rtle},\ and\ \citenamefont
  {Thoss}}]{wang2011numerically}%
  \BibitemOpen
  \bibfield  {author} {\bibinfo {author} {\bibfnamefont {H.}~\bibnamefont
  {Wang}}, \bibinfo {author} {\bibfnamefont {I.}~\bibnamefont {Pshenichnyuk}},
  \bibinfo {author} {\bibfnamefont {R.}~\bibnamefont {H{\"a}rtle}},\ and\
  \bibinfo {author} {\bibfnamefont {M.}~\bibnamefont {Thoss}},\ }\bibfield
  {title} {\enquote {\bibinfo {title} {Numerically exact, time-dependent
  treatment of vibrationally coupled electron transport in single-molecule
  junctions},}\ }\href@noop {} {\bibfield  {journal} {\bibinfo  {journal} {The
  Journal of chemical physics}\ }\textbf {\bibinfo {volume} {135}} (\bibinfo
  {year} {2011})}\BibitemShut {NoStop}%
\bibitem [{\citenamefont {Wang}\ and\ \citenamefont
  {Thoss}(2008)}]{wang2008coherent}%
  \BibitemOpen
  \bibfield  {author} {\bibinfo {author} {\bibfnamefont {H.}~\bibnamefont
  {Wang}}\ and\ \bibinfo {author} {\bibfnamefont {M.}~\bibnamefont {Thoss}},\
  }\bibfield  {title} {\enquote {\bibinfo {title} {From coherent motion to
  localization: dynamics of the spin-boson model at zero temperature},}\
  }\href@noop {} {\bibfield  {journal} {\bibinfo  {journal} {New Journal of
  Physics}\ }\textbf {\bibinfo {volume} {10}},\ \bibinfo {pages} {115005}
  (\bibinfo {year} {2008})}\BibitemShut {NoStop}%
\bibitem [{\citenamefont {Matzkies}\ and\ \citenamefont
  {Manthe}(1998)}]{matzkies1998accurate}%
  \BibitemOpen
  \bibfield  {author} {\bibinfo {author} {\bibfnamefont {F.}~\bibnamefont
  {Matzkies}}\ and\ \bibinfo {author} {\bibfnamefont {U.}~\bibnamefont
  {Manthe}},\ }\bibfield  {title} {\enquote {\bibinfo {title} {Accurate quantum
  calculations of thermal rate constants employing mctdh: H 2+ oh→ h+ h 2 o
  and d 2+ oh→ d+ doh},}\ }\href@noop {} {\bibfield  {journal} {\bibinfo
  {journal} {The Journal of chemical physics}\ }\textbf {\bibinfo {volume}
  {108}},\ \bibinfo {pages} {4828--4836} (\bibinfo {year} {1998})}\BibitemShut
  {NoStop}%
\bibitem [{\citenamefont {Huarte-Larranaga}\ and\ \citenamefont
  {Manthe}(2000)}]{huarte2000full}%
  \BibitemOpen
  \bibfield  {author} {\bibinfo {author} {\bibfnamefont {F.}~\bibnamefont
  {Huarte-Larranaga}}\ and\ \bibinfo {author} {\bibfnamefont {U.}~\bibnamefont
  {Manthe}},\ }\bibfield  {title} {\enquote {\bibinfo {title} {Full dimensional
  quantum calculations of the ch 4+ h→ ch 3+ h 2 reaction rate},}\
  }\href@noop {} {\bibfield  {journal} {\bibinfo  {journal} {Journal of
  Chemical Physics}\ }\textbf {\bibinfo {volume} {113}},\ \bibinfo {pages}
  {5115--5118} (\bibinfo {year} {2000})}\BibitemShut {NoStop}%
\bibitem [{\citenamefont {Mellini}\ and\ \citenamefont
  {Vendrell}(2025)}]{mellini2025competition}%
  \BibitemOpen
  \bibfield  {author} {\bibinfo {author} {\bibfnamefont {F.}~\bibnamefont
  {Mellini}}\ and\ \bibinfo {author} {\bibfnamefont {O.}~\bibnamefont
  {Vendrell}},\ }\bibfield  {title} {\enquote {\bibinfo {title} {Competition
  between coherent ultrafast energy redistribution and photochemistry in the
  collective strong coupling regime: The role of static disorder},}\
  }\href@noop {} {\bibfield  {journal} {\bibinfo  {journal} {The Journal of
  Physical Chemistry Letters}\ }\textbf {\bibinfo {volume} {16}},\ \bibinfo
  {pages} {6155--6162} (\bibinfo {year} {2025})}\BibitemShut {NoStop}%
\bibitem [{\citenamefont {Wallner}, \citenamefont {Remnant},\ and\
  \citenamefont {Vendrell}(2024)}]{wallner2024strong}%
  \BibitemOpen
  \bibfield  {author} {\bibinfo {author} {\bibfnamefont {L.}~\bibnamefont
  {Wallner}}, \bibinfo {author} {\bibfnamefont {C.}~\bibnamefont {Remnant}},\
  and\ \bibinfo {author} {\bibfnamefont {O.}~\bibnamefont {Vendrell}},\
  }\bibfield  {title} {\enquote {\bibinfo {title} {Strong-coupling modification
  of singlet-fission dynamical pathways},}\ }\href@noop {} {\bibfield
  {journal} {\bibinfo  {journal} {The Journal of Physical Chemistry A}\
  }\textbf {\bibinfo {volume} {128}},\ \bibinfo {pages} {8897--8905} (\bibinfo
  {year} {2024})}\BibitemShut {NoStop}%
\bibitem [{\citenamefont {Krupp}, \citenamefont {Groenhof},\ and\ \citenamefont
  {Vendrell}(2025)}]{krupp2025quantum}%
  \BibitemOpen
  \bibfield  {author} {\bibinfo {author} {\bibfnamefont {N.}~\bibnamefont
  {Krupp}}, \bibinfo {author} {\bibfnamefont {G.}~\bibnamefont {Groenhof}},\
  and\ \bibinfo {author} {\bibfnamefont {O.}~\bibnamefont {Vendrell}},\
  }\bibfield  {title} {\enquote {\bibinfo {title} {Quantum dynamics simulation
  of exciton-polariton transport},}\ }\href@noop {} {\bibfield  {journal}
  {\bibinfo  {journal} {Nature Communications}\ }\textbf {\bibinfo {volume}
  {16}},\ \bibinfo {pages} {5431} (\bibinfo {year} {2025})}\BibitemShut
  {NoStop}%
\bibitem [{\citenamefont {Raab}, \citenamefont {Burghardt},\ and\ \citenamefont
  {Meyer}(1999)}]{raab_multiconfiguration_1999}%
  \BibitemOpen
  \bibfield  {author} {\bibinfo {author} {\bibfnamefont {A.}~\bibnamefont
  {Raab}}, \bibinfo {author} {\bibfnamefont {I.}~\bibnamefont {Burghardt}},\
  and\ \bibinfo {author} {\bibfnamefont {H.-D.}\ \bibnamefont {Meyer}},\
  }\bibfield  {title} {\enquote {\bibinfo {title} {The multiconfiguration
  time-dependent {Hartree} method generalized to the propagation of density
  operators},}\ }\href {https://doi.org/10.1063/1.480334} {\bibfield  {journal}
  {\bibinfo  {journal} {The Journal of Chemical Physics}\ }\textbf {\bibinfo
  {volume} {111}},\ \bibinfo {pages} {8759--8772} (\bibinfo {year}
  {1999})}\BibitemShut {NoStop}%
\bibitem [{\citenamefont {Raab}\ and\ \citenamefont
  {Meyer}(2000)}]{raab_multiconfigurational_2000}%
  \BibitemOpen
  \bibfield  {author} {\bibinfo {author} {\bibfnamefont {A.}~\bibnamefont
  {Raab}}\ and\ \bibinfo {author} {\bibfnamefont {H.-D.}\ \bibnamefont
  {Meyer}},\ }\bibfield  {title} {\enquote {\bibinfo {title}
  {Multiconfigurational expansions of density operators: equations of motion
  and their properties},}\ }\href {https://doi.org/10.1007/s002140000146}
  {\bibfield  {journal} {\bibinfo  {journal} {Theoretical Chemistry Accounts:
  Theory, Computation, and Modeling (Theoretica Chimica Acta)}\ }\textbf
  {\bibinfo {volume} {104}},\ \bibinfo {pages} {358--369} (\bibinfo {year}
  {2000})}\BibitemShut {NoStop}%
\bibitem [{\citenamefont {Van~Haeften}, \citenamefont {Ash},\ and\
  \citenamefont {Worth}(2023)}]{van_haeften_propagating_2023}%
  \BibitemOpen
  \bibfield  {author} {\bibinfo {author} {\bibfnamefont {A.}~\bibnamefont
  {Van~Haeften}}, \bibinfo {author} {\bibfnamefont {C.}~\bibnamefont {Ash}},\
  and\ \bibinfo {author} {\bibfnamefont {G.}~\bibnamefont {Worth}},\ }\bibfield
   {title} {\enquote {\bibinfo {title} {Propagating multi-dimensional density
  operators using the multi-layer- \textit{$\rho$} multi-configurational
  time-dependent {Hartree} method},}\ }\href
  {https://doi.org/10.1063/5.0172956} {\bibfield  {journal} {\bibinfo
  {journal} {The Journal of Chemical Physics}\ }\textbf {\bibinfo {volume}
  {159}},\ \bibinfo {pages} {194114} (\bibinfo {year} {2023})}\BibitemShut
  {NoStop}%
\bibitem [{\citenamefont {Gelman}\ and\ \citenamefont
  {Kosloff}(2003)}]{gelman_simulating_2003}%
  \BibitemOpen
  \bibfield  {author} {\bibinfo {author} {\bibfnamefont {D.}~\bibnamefont
  {Gelman}}\ and\ \bibinfo {author} {\bibfnamefont {R.}~\bibnamefont
  {Kosloff}},\ }\bibfield  {title} {\enquote {\bibinfo {title} {Simulating
  dissipative phenomena with a random phase thermal wavefunctions, high
  temperature application of the {Surrogate} {Hamiltonian} approach},}\ }\href
  {https://doi.org/10.1016/j.cplett.2003.09.119} {\bibfield  {journal}
  {\bibinfo  {journal} {Chemical Physics Letters}\ }\textbf {\bibinfo {volume}
  {381}},\ \bibinfo {pages} {129--138} (\bibinfo {year} {2003})}\BibitemShut
  {NoStop}%
\bibitem [{\citenamefont {Nest}\ and\ \citenamefont
  {Kosloff}(2007)}]{nest_quantum_2007}%
  \BibitemOpen
  \bibfield  {author} {\bibinfo {author} {\bibfnamefont {M.}~\bibnamefont
  {Nest}}\ and\ \bibinfo {author} {\bibfnamefont {R.}~\bibnamefont {Kosloff}},\
  }\bibfield  {title} {\enquote {\bibinfo {title} {Quantum dynamical treatment
  of inelastic scattering of atoms at a surface at finite temperature: {The}
  random phase thermal wave function approach},}\ }\href
  {https://doi.org/10.1063/1.2786088} {\bibfield  {journal} {\bibinfo
  {journal} {The Journal of Chemical Physics}\ }\textbf {\bibinfo {volume}
  {127}},\ \bibinfo {pages} {134711} (\bibinfo {year} {2007})}\BibitemShut
  {NoStop}%
\bibitem [{\citenamefont {Wang}, \citenamefont {Skinner},\ and\ \citenamefont
  {Thoss}(2006)}]{wang_calculation_2006}%
  \BibitemOpen
  \bibfield  {author} {\bibinfo {author} {\bibfnamefont {H.}~\bibnamefont
  {Wang}}, \bibinfo {author} {\bibfnamefont {D.~E.}\ \bibnamefont {Skinner}},\
  and\ \bibinfo {author} {\bibfnamefont {M.}~\bibnamefont {Thoss}},\ }\bibfield
   {title} {\enquote {\bibinfo {title} {Calculation of reactive flux
  correlation functions for systems in a condensed phase environment: {A}
  multilayer multiconfiguration time-dependent {Hartree} approach},}\ }\href
  {https://doi.org/10.1063/1.2363195} {\bibfield  {journal} {\bibinfo
  {journal} {The Journal of Chemical Physics}\ }\textbf {\bibinfo {volume}
  {125}},\ \bibinfo {pages} {174502} (\bibinfo {year} {2006})}\BibitemShut
  {NoStop}%
\bibitem [{\citenamefont {Wang}\ and\ \citenamefont
  {Thoss}(2006)}]{wang_quantum-mechanical_2006}%
  \BibitemOpen
  \bibfield  {author} {\bibinfo {author} {\bibfnamefont {H.}~\bibnamefont
  {Wang}}\ and\ \bibinfo {author} {\bibfnamefont {M.}~\bibnamefont {Thoss}},\
  }\bibfield  {title} {\enquote {\bibinfo {title} {Quantum-mechanical
  evaluation of the {Boltzmann} operator in correlation functions for large
  molecular systems: {A} multilayer multiconfiguration time-dependent {Hartree}
  approach},}\ }\href {https://doi.org/10.1063/1.2161178} {\bibfield  {journal}
  {\bibinfo  {journal} {The Journal of Chemical Physics}\ }\textbf {\bibinfo
  {volume} {124}},\ \bibinfo {pages} {034114} (\bibinfo {year}
  {2006})}\BibitemShut {NoStop}%
\bibitem [{\citenamefont {Barnett}\ and\ \citenamefont
  {Dalton}(1987)}]{barnett_liouville_1987}%
  \BibitemOpen
  \bibfield  {author} {\bibinfo {author} {\bibfnamefont {S.~M.}\ \bibnamefont
  {Barnett}}\ and\ \bibinfo {author} {\bibfnamefont {B.~J.}\ \bibnamefont
  {Dalton}},\ }\bibfield  {title} {\enquote {\bibinfo {title} {Liouville space
  description of thermofields and their generalisations},}\ }\href
  {https://doi.org/10.1088/0305-4470/20/2/026} {\bibfield  {journal} {\bibinfo
  {journal} {J. Phys. A: Math. Gen.}\ }\textbf {\bibinfo {volume} {20}},\
  \bibinfo {pages} {411--418} (\bibinfo {year} {1987})}\BibitemShut {NoStop}%
\bibitem [{\citenamefont {Borrelli}\ and\ \citenamefont
  {Gelin}(2016)}]{borrelli_quantum_2016}%
  \BibitemOpen
  \bibfield  {author} {\bibinfo {author} {\bibfnamefont {R.}~\bibnamefont
  {Borrelli}}\ and\ \bibinfo {author} {\bibfnamefont {M.~F.}\ \bibnamefont
  {Gelin}},\ }\bibfield  {title} {\enquote {\bibinfo {title} {Quantum
  electron-vibrational dynamics at finite temperature: {Thermo} field dynamics
  approach},}\ }\href {https://doi.org/10.1063/1.4971211} {\bibfield  {journal}
  {\bibinfo  {journal} {The Journal of Chemical Physics}\ }\textbf {\bibinfo
  {volume} {145}},\ \bibinfo {pages} {224101} (\bibinfo {year}
  {2016})}\BibitemShut {NoStop}%
\bibitem [{\citenamefont {Fischer}\ and\ \citenamefont
  {Saalfrank}(2021)}]{fischer_thermofield-based_2021}%
  \BibitemOpen
  \bibfield  {author} {\bibinfo {author} {\bibfnamefont {E.~W.}\ \bibnamefont
  {Fischer}}\ and\ \bibinfo {author} {\bibfnamefont {P.}~\bibnamefont
  {Saalfrank}},\ }\bibfield  {title} {\enquote {\bibinfo {title} {A
  thermofield-based multilayer multiconfigurational time-dependent {Hartree}
  approach to non-adiabatic quantum dynamics at finite temperature},}\ }\href
  {https://doi.org/10.1063/5.0064013} {\bibfield  {journal} {\bibinfo
  {journal} {The Journal of Chemical Physics}\ }\textbf {\bibinfo {volume}
  {155}},\ \bibinfo {pages} {134109} (\bibinfo {year} {2021})}\BibitemShut
  {NoStop}%
\bibitem [{\citenamefont {Brey}\ \emph {et~al.}(2021)\citenamefont {Brey},
  \citenamefont {Popp}, \citenamefont {Budakoti}, \citenamefont {D’Avino},\
  and\ \citenamefont {Burghardt}}]{brey_quantum_2021}%
  \BibitemOpen
  \bibfield  {author} {\bibinfo {author} {\bibfnamefont {D.}~\bibnamefont
  {Brey}}, \bibinfo {author} {\bibfnamefont {W.}~\bibnamefont {Popp}}, \bibinfo
  {author} {\bibfnamefont {P.}~\bibnamefont {Budakoti}}, \bibinfo {author}
  {\bibfnamefont {G.}~\bibnamefont {D’Avino}},\ and\ \bibinfo {author}
  {\bibfnamefont {I.}~\bibnamefont {Burghardt}},\ }\bibfield  {title} {\enquote
  {\bibinfo {title} {Quantum {Dynamics} of {Electron}–{Hole} {Separation} in
  {Stacked} {Perylene} {Diimide}-{Based} {Self}-{Assembled}
  {Nanostructures}},}\ }\href {https://doi.org/10.1021/acs.jpcc.1c06374}
  {\bibfield  {journal} {\bibinfo  {journal} {J. Phys. Chem. C}\ }\textbf
  {\bibinfo {volume} {125}},\ \bibinfo {pages} {25030--25043} (\bibinfo {year}
  {2021})}\BibitemShut {NoStop}%
\bibitem [{\citenamefont {Verstraete}, \citenamefont {García-Ripoll},\ and\
  \citenamefont {Cirac}(2004)}]{verstraete_matrix_2004}%
  \BibitemOpen
  \bibfield  {author} {\bibinfo {author} {\bibfnamefont {F.}~\bibnamefont
  {Verstraete}}, \bibinfo {author} {\bibfnamefont {J.~J.}\ \bibnamefont
  {García-Ripoll}},\ and\ \bibinfo {author} {\bibfnamefont {J.~I.}\
  \bibnamefont {Cirac}},\ }\bibfield  {title} {\enquote {\bibinfo {title}
  {Matrix {Product} {Density} {Operators}: {Simulation} of
  {Finite}-{Temperature} and {Dissipative} {Systems}},}\ }\href
  {https://doi.org/10.1103/PhysRevLett.93.207204} {\bibfield  {journal}
  {\bibinfo  {journal} {Phys. Rev. Lett.}\ }\textbf {\bibinfo {volume} {93}},\
  \bibinfo {pages} {207204} (\bibinfo {year} {2004})}\BibitemShut {NoStop}%
\bibitem [{\citenamefont {Zwolak}\ and\ \citenamefont
  {Vidal}(2004)}]{zwolak2004mixed}%
  \BibitemOpen
  \bibfield  {author} {\bibinfo {author} {\bibfnamefont {M.}~\bibnamefont
  {Zwolak}}\ and\ \bibinfo {author} {\bibfnamefont {G.}~\bibnamefont {Vidal}},\
  }\bibfield  {title} {\enquote {\bibinfo {title} {Mixed-state dynamics in
  one-dimensional quantum lattice systems: Time-dependent superoperator
  renormalization algorithm},}\ }\href@noop {} {\bibfield  {journal} {\bibinfo
  {journal} {Physical review letters}\ }\textbf {\bibinfo {volume} {93}},\
  \bibinfo {pages} {207205} (\bibinfo {year} {2004})}\BibitemShut {NoStop}%
\bibitem [{\citenamefont {White}(2009)}]{white_minimally_2009}%
  \BibitemOpen
  \bibfield  {author} {\bibinfo {author} {\bibfnamefont {S.~R.}\ \bibnamefont
  {White}},\ }\bibfield  {title} {\enquote {\bibinfo {title} {Minimally
  {Entangled} {Typical} {Quantum} {States} at {Finite} {Temperature}},}\ }\href
  {https://doi.org/10.1103/PhysRevLett.102.190601} {\bibfield  {journal}
  {\bibinfo  {journal} {Phys. Rev. Lett.}\ }\textbf {\bibinfo {volume} {102}},\
  \bibinfo {pages} {190601} (\bibinfo {year} {2009})}\BibitemShut {NoStop}%
\bibitem [{\citenamefont {Chen}\ and\ \citenamefont
  {Stoudenmire}(2020)}]{chen_hybrid_2020}%
  \BibitemOpen
  \bibfield  {author} {\bibinfo {author} {\bibfnamefont {J.}~\bibnamefont
  {Chen}}\ and\ \bibinfo {author} {\bibfnamefont {E.~M.}\ \bibnamefont
  {Stoudenmire}},\ }\bibfield  {title} {\enquote {\bibinfo {title} {Hybrid
  purification and sampling approach for thermal quantum systems},}\ }\href
  {https://doi.org/10.1103/PhysRevB.101.195119} {\bibfield  {journal} {\bibinfo
   {journal} {Phys. Rev. B}\ }\textbf {\bibinfo {volume} {101}},\ \bibinfo
  {pages} {195119} (\bibinfo {year} {2020})}\BibitemShut {NoStop}%
\bibitem [{\citenamefont {Feiguin}\ and\ \citenamefont
  {White}(2005)}]{feiguin_finite-temperature_2005}%
  \BibitemOpen
  \bibfield  {author} {\bibinfo {author} {\bibfnamefont {A.~E.}\ \bibnamefont
  {Feiguin}}\ and\ \bibinfo {author} {\bibfnamefont {S.~R.}\ \bibnamefont
  {White}},\ }\bibfield  {title} {\enquote {\bibinfo {title}
  {Finite-temperature density matrix renormalization using an enlarged
  {Hilbert} space},}\ }\href {https://doi.org/10.1103/PhysRevB.72.220401}
  {\bibfield  {journal} {\bibinfo  {journal} {Phys. Rev. B}\ }\textbf {\bibinfo
  {volume} {72}},\ \bibinfo {pages} {220401} (\bibinfo {year}
  {2005})}\BibitemShut {NoStop}%
\bibitem [{\citenamefont {Verstraete}, \citenamefont {Porras},\ and\
  \citenamefont {Cirac}(2004)}]{verstraete2004density}%
  \BibitemOpen
  \bibfield  {author} {\bibinfo {author} {\bibfnamefont {F.}~\bibnamefont
  {Verstraete}}, \bibinfo {author} {\bibfnamefont {D.}~\bibnamefont {Porras}},\
  and\ \bibinfo {author} {\bibfnamefont {J.~I.}\ \bibnamefont {Cirac}},\
  }\bibfield  {title} {\enquote {\bibinfo {title} {Density matrix
  renormalization group and periodic boundary conditions: A quantum information
  perspective},}\ }\href@noop {} {\bibfield  {journal} {\bibinfo  {journal}
  {Physical review letters}\ }\textbf {\bibinfo {volume} {93}},\ \bibinfo
  {pages} {227205} (\bibinfo {year} {2004})}\BibitemShut {NoStop}%
\bibitem [{\citenamefont {Tannor}(2008)}]{tannor2008introduction}%
  \BibitemOpen
  \bibfield  {author} {\bibinfo {author} {\bibfnamefont {D.}~\bibnamefont
  {Tannor}},\ }\href@noop {} {\emph {\bibinfo {title} {Introduction to quantum
  mechanics: a time-dependent perspective}}}\ (\bibinfo  {publisher} {MIT
  Press},\ \bibinfo {year} {2008})\BibitemShut {NoStop}%
\bibitem [{\citenamefont {Arimitsu}\ and\ \citenamefont
  {Umezawa}(1985)}]{ari85:429}%
  \BibitemOpen
  \bibfield  {author} {\bibinfo {author} {\bibfnamefont {T.}~\bibnamefont
  {Arimitsu}}\ and\ \bibinfo {author} {\bibfnamefont {H.}~\bibnamefont
  {Umezawa}},\ }\bibfield  {title} {\enquote {\bibinfo {title} {A {{General
  Formulation}} of {{Nonequilibrium Thermo Field Dynamics}}},}\ }\href
  {https://doi.org/10.1143/PTP.74.429} {\bibfield  {journal} {\bibinfo
  {journal} {Progress of Theoretical Physics}\ }\textbf {\bibinfo {volume}
  {74}},\ \bibinfo {pages} {429--432} (\bibinfo {year} {1985})}\BibitemShut
  {NoStop}%
\bibitem [{\citenamefont {Barnett}\ and\ \citenamefont
  {Knight}(1985)}]{bar85:467}%
  \BibitemOpen
  \bibfield  {author} {\bibinfo {author} {\bibfnamefont {S.~M.}\ \bibnamefont
  {Barnett}}\ and\ \bibinfo {author} {\bibfnamefont {P.~L.}\ \bibnamefont
  {Knight}},\ }\bibfield  {title} {\enquote {\bibinfo {title} {Thermofield
  analysis of squeezing and statistical mixtures in quantum optics},}\ }\href
  {https://doi.org/10.1364/JOSAB.2.000467} {\bibfield  {journal} {\bibinfo
  {journal} {Journal of the Optical Society of America B}\ }\textbf {\bibinfo
  {volume} {2}},\ \bibinfo {pages} {467} (\bibinfo {year} {1985})}\BibitemShut
  {NoStop}%
\bibitem [{\citenamefont {Harsha}, \citenamefont {Henderson},\ and\
  \citenamefont {Scuseria}(2019)}]{harsha_thermofield_2019}%
  \BibitemOpen
  \bibfield  {author} {\bibinfo {author} {\bibfnamefont {G.}~\bibnamefont
  {Harsha}}, \bibinfo {author} {\bibfnamefont {T.~M.}\ \bibnamefont
  {Henderson}},\ and\ \bibinfo {author} {\bibfnamefont {G.~E.}\ \bibnamefont
  {Scuseria}},\ }\bibfield  {title} {\enquote {\bibinfo {title} {Thermofield
  theory for finite-temperature quantum chemistry},}\ }\href
  {https://doi.org/10.1063/1.5089560} {\bibfield  {journal} {\bibinfo
  {journal} {The Journal of Chemical Physics}\ }\textbf {\bibinfo {volume}
  {150}},\ \bibinfo {pages} {154109} (\bibinfo {year} {2019})}\BibitemShut
  {NoStop}%
\bibitem [{\citenamefont {Borrelli}\ and\ \citenamefont
  {Gelin}(2021)}]{borrelli_finite_2021}%
  \BibitemOpen
  \bibfield  {author} {\bibinfo {author} {\bibfnamefont {R.}~\bibnamefont
  {Borrelli}}\ and\ \bibinfo {author} {\bibfnamefont {M.~F.}\ \bibnamefont
  {Gelin}},\ }\bibfield  {title} {\enquote {\bibinfo {title} {Finite
  temperature quantum dynamics of complex systems: {Integrating}
  {\textless}span
  style="font-variant:small-caps;"{\textgreater}thermo‐field{\textless}/span{\textgreater}
  theories and {\textless}span
  style="font-variant:small-caps;"{\textgreater}tensor‐train{\textless}/span{\textgreater}
  methods},}\ }\href {https://doi.org/10.1002/wcms.1539} {\bibfield  {journal}
  {\bibinfo  {journal} {WIREs Comput Mol Sci}\ }\textbf {\bibinfo {volume}
  {11}},\ \bibinfo {pages} {e1539} (\bibinfo {year} {2021})}\BibitemShut
  {NoStop}%
\bibitem [{\citenamefont {Shushkov}\ and\ \citenamefont
  {Miller~III}(2019)}]{shu19:134107}%
  \BibitemOpen
  \bibfield  {author} {\bibinfo {author} {\bibfnamefont {P.}~\bibnamefont
  {Shushkov}}\ and\ \bibinfo {author} {\bibfnamefont {T.~F.}\ \bibnamefont
  {Miller~III}},\ }\bibfield  {title} {\enquote {\bibinfo {title} {Real-time
  density-matrix coupled-cluster approach for closed and open systems at finite
  temperature},}\ }\href {https://doi.org/10.1063/1.5121749} {\bibfield
  {journal} {\bibinfo  {journal} {The Journal of Chemical Physics}\ }\textbf
  {\bibinfo {volume} {151}},\ \bibinfo {pages} {134107} (\bibinfo {year}
  {2019})},\ \Eprint {https://arxiv.org/abs/1907.11962} {arXiv:1907.11962}
  \BibitemShut {NoStop}%
\bibitem [{\citenamefont {Larsson}(2024)}]{lar24:e2306881}%
  \BibitemOpen
  \bibfield  {author} {\bibinfo {author} {\bibfnamefont {H.~R.}\ \bibnamefont
  {Larsson}},\ }\bibfield  {title} {\enquote {\bibinfo {title} {A tensor
  network view of multilayer multiconfiguration time-dependent {{Hartree}}
  methods},}\ }\href {https://doi.org/10.1080/00268976.2024.2306881} {\bibfield
   {journal} {\bibinfo  {journal} {Molecular Physics}\ }\textbf {\bibinfo
  {volume} {122}},\ \bibinfo {pages} {e2306881} (\bibinfo {year}
  {2024})}\BibitemShut {NoStop}%
\bibitem [{\citenamefont {Weike}\ and\ \citenamefont
  {Manthe}(2021)}]{wei21:194108}%
  \BibitemOpen
  \bibfield  {author} {\bibinfo {author} {\bibfnamefont {T.}~\bibnamefont
  {Weike}}\ and\ \bibinfo {author} {\bibfnamefont {U.}~\bibnamefont {Manthe}},\
  }\bibfield  {title} {\enquote {\bibinfo {title} {Symmetries in the
  multi-configurational time-dependent {{Hartree}} wavefunction representation
  and propagation},}\ }\href {https://doi.org/10.1063/5.0054105} {\bibfield
  {journal} {\bibinfo  {journal} {The Journal of Chemical Physics}\ }\textbf
  {\bibinfo {volume} {154}},\ \bibinfo {pages} {194108} (\bibinfo {year}
  {2021})}\BibitemShut {NoStop}%
\bibitem [{\citenamefont {Manthe}\ and\ \citenamefont
  {Huarte-Larrañaga}(2001)}]{manthe_partition_2001}%
  \BibitemOpen
  \bibfield  {author} {\bibinfo {author} {\bibfnamefont {U.}~\bibnamefont
  {Manthe}}\ and\ \bibinfo {author} {\bibfnamefont {F.}~\bibnamefont
  {Huarte-Larrañaga}},\ }\bibfield  {title} {\enquote {\bibinfo {title}
  {Partition functions for reaction rate calculations: statistical sampling and
  {MCTDH} propagation},}\ }\href
  {https://doi.org/10.1016/S0009-2614(01)01207-6} {\bibfield  {journal}
  {\bibinfo  {journal} {Chemical Physics Letters}\ }\textbf {\bibinfo {volume}
  {349}},\ \bibinfo {pages} {321--328} (\bibinfo {year} {2001})}\BibitemShut
  {NoStop}%
\bibitem [{\citenamefont {Worth}\ \emph {et~al.}()\citenamefont {Worth},
  \citenamefont {Beck}, \citenamefont {J{\"a}ckle}, \citenamefont {Vendrell},\
  and\ \citenamefont {Meyer}}]{mctdh:MLpackage}%
  \BibitemOpen
  \bibfield  {author} {\bibinfo {author} {\bibfnamefont {G.~A.}\ \bibnamefont
  {Worth}}, \bibinfo {author} {\bibfnamefont {M.~H.}\ \bibnamefont {Beck}},
  \bibinfo {author} {\bibfnamefont {A.}~\bibnamefont {J{\"a}ckle}}, \bibinfo
  {author} {\bibfnamefont {O.}~\bibnamefont {Vendrell}},\ and\ \bibinfo
  {author} {\bibfnamefont {H.-D.}\ \bibnamefont {Meyer}},\ }\href@noop {}
  {}\bibinfo {howpublished} {{The MCTDH Package, Version 8.2, (2000). H.-D.
  Meyer, Version 8.3 (2002), {V}ersion 8.4 (2007). O. Vendrell and H.-D. Meyer
  {V}ersion 8.5 (2013). Versions 8.5 and 8.6 contains the ML-MCTDH algorithm.
  Used version: 8.6.10 (March 2026). {S}ee
  http://mctdh.uni-hd.de/}}\BibitemShut {NoStop}%
\bibitem [{\citenamefont {Polyansky}, \citenamefont {Jensen},\ and\
  \citenamefont {Tennyson}(1994)}]{polyansky1994spectroscopically}%
  \BibitemOpen
  \bibfield  {author} {\bibinfo {author} {\bibfnamefont {O.~L.}\ \bibnamefont
  {Polyansky}}, \bibinfo {author} {\bibfnamefont {P.}~\bibnamefont {Jensen}},\
  and\ \bibinfo {author} {\bibfnamefont {J.}~\bibnamefont {Tennyson}},\
  }\bibfield  {title} {\enquote {\bibinfo {title} {A spectroscopically
  determined potential energy surface for the ground state of h216o: A new
  level of accuracy},}\ }\href@noop {} {\bibfield  {journal} {\bibinfo
  {journal} {The Journal of chemical physics}\ }\textbf {\bibinfo {volume}
  {101}},\ \bibinfo {pages} {7651--7657} (\bibinfo {year} {1994})}\BibitemShut
  {NoStop}%
\bibitem [{\citenamefont {Polyansky}, \citenamefont {Jensen},\ and\
  \citenamefont {Tennyson}(1996)}]{polyansky1996potential}%
  \BibitemOpen
  \bibfield  {author} {\bibinfo {author} {\bibfnamefont {O.~L.}\ \bibnamefont
  {Polyansky}}, \bibinfo {author} {\bibfnamefont {P.}~\bibnamefont {Jensen}},\
  and\ \bibinfo {author} {\bibfnamefont {J.}~\bibnamefont {Tennyson}},\
  }\bibfield  {title} {\enquote {\bibinfo {title} {The potential energy surface
  of h2 16o},}\ }\href@noop {} {\bibfield  {journal} {\bibinfo  {journal} {The
  Journal of chemical physics}\ }\textbf {\bibinfo {volume} {105}},\ \bibinfo
  {pages} {6490--6497} (\bibinfo {year} {1996})}\BibitemShut {NoStop}%
\bibitem [{\citenamefont {Pel{\'a}ez}, \citenamefont {Sadri},\ and\
  \citenamefont {Meyer}(2014)}]{pelaez2014full}%
  \BibitemOpen
  \bibfield  {author} {\bibinfo {author} {\bibfnamefont {D.}~\bibnamefont
  {Pel{\'a}ez}}, \bibinfo {author} {\bibfnamefont {K.}~\bibnamefont {Sadri}},\
  and\ \bibinfo {author} {\bibfnamefont {H.-D.}\ \bibnamefont {Meyer}},\
  }\bibfield  {title} {\enquote {\bibinfo {title} {Full-dimensional mctdh/mgpf
  study of the ground and lowest lying vibrational states of the bihydroxide
  h3o2-complex},}\ }\href@noop {} {\bibfield  {journal} {\bibinfo  {journal}
  {Spectrochimica Acta Part A: Molecular and Biomolecular Spectroscopy}\
  }\textbf {\bibinfo {volume} {119}},\ \bibinfo {pages} {42--51} (\bibinfo
  {year} {2014})}\BibitemShut {NoStop}%
\bibitem [{\citenamefont {Pel{\'a}ez}\ and\ \citenamefont
  {Meyer}(2017)}]{pelaez2017infrared}%
  \BibitemOpen
  \bibfield  {author} {\bibinfo {author} {\bibfnamefont {D.}~\bibnamefont
  {Pel{\'a}ez}}\ and\ \bibinfo {author} {\bibfnamefont {H.-D.}\ \bibnamefont
  {Meyer}},\ }\bibfield  {title} {\enquote {\bibinfo {title} {On the infrared
  absorption spectrum of the hydrated hydroxide (h3o2-) cluster anion},}\
  }\href@noop {} {\bibfield  {journal} {\bibinfo  {journal} {Chemical Physics}\
  }\textbf {\bibinfo {volume} {482}},\ \bibinfo {pages} {100--105} (\bibinfo
  {year} {2017})}\BibitemShut {NoStop}%
\bibitem [{\citenamefont {Vendrell}, \citenamefont {Gatti},\ and\ \citenamefont
  {Meyer}(2007)}]{vendrell2007full}%
  \BibitemOpen
  \bibfield  {author} {\bibinfo {author} {\bibfnamefont {O.}~\bibnamefont
  {Vendrell}}, \bibinfo {author} {\bibfnamefont {F.}~\bibnamefont {Gatti}},\
  and\ \bibinfo {author} {\bibfnamefont {H.-D.}\ \bibnamefont {Meyer}},\
  }\bibfield  {title} {\enquote {\bibinfo {title} {Full dimensional
  (15-dimensional) quantum-dynamical simulation of the protonated water dimer.
  ii. infrared spectrum and vibrational dynamics},}\ }\href@noop {} {\bibfield
  {journal} {\bibinfo  {journal} {The Journal of chemical physics}\ }\textbf
  {\bibinfo {volume} {127}} (\bibinfo {year} {2007})}\BibitemShut {NoStop}%
\bibitem [{\citenamefont {Bunker}\ and\ \citenamefont
  {Jensen}(2018)}]{bunker2018fundamentals}%
  \BibitemOpen
  \bibfield  {author} {\bibinfo {author} {\bibfnamefont {P.~R.}\ \bibnamefont
  {Bunker}}\ and\ \bibinfo {author} {\bibfnamefont {P.}~\bibnamefont
  {Jensen}},\ }\href@noop {} {\emph {\bibinfo {title} {Fundamentals of
  molecular symmetry}}}\ (\bibinfo  {publisher} {CRC Press},\ \bibinfo {year}
  {2018})\BibitemShut {NoStop}%
\bibitem [{\citenamefont {Beyer}\ and\ \citenamefont
  {Swinehart}(1973)}]{beyer1973algorithm}%
  \BibitemOpen
  \bibfield  {author} {\bibinfo {author} {\bibfnamefont {T.}~\bibnamefont
  {Beyer}}\ and\ \bibinfo {author} {\bibfnamefont {D.}~\bibnamefont
  {Swinehart}},\ }\bibfield  {title} {\enquote {\bibinfo {title} {Algorithm
  448: number of multiply-restricted partitions},}\ }\href@noop {} {\bibfield
  {journal} {\bibinfo  {journal} {Communications of the ACM}\ }\textbf
  {\bibinfo {volume} {16}},\ \bibinfo {pages} {379} (\bibinfo {year}
  {1973})}\BibitemShut {NoStop}%
\bibitem [{\citenamefont {Stein}\ and\ \citenamefont
  {Rabinovitch}(1973)}]{stein1973accurate}%
  \BibitemOpen
  \bibfield  {author} {\bibinfo {author} {\bibfnamefont {S.~E.}\ \bibnamefont
  {Stein}}\ and\ \bibinfo {author} {\bibfnamefont {B.}~\bibnamefont
  {Rabinovitch}},\ }\bibfield  {title} {\enquote {\bibinfo {title} {Accurate
  evaluation of internal energy level sums and densities including anharmonic
  oscillators and hindered rotors},}\ }\href@noop {} {\bibfield  {journal}
  {\bibinfo  {journal} {The Journal of Chemical Physics}\ }\textbf {\bibinfo
  {volume} {58}},\ \bibinfo {pages} {2438--2445} (\bibinfo {year}
  {1973})}\BibitemShut {NoStop}%
\bibitem [{\citenamefont {Mendive-Tapia}, \citenamefont {Meyer},\ and\
  \citenamefont {Vendrell}(2023)}]{mendive2023optimal}%
  \BibitemOpen
  \bibfield  {author} {\bibinfo {author} {\bibfnamefont {D.}~\bibnamefont
  {Mendive-Tapia}}, \bibinfo {author} {\bibfnamefont {H.-D.}\ \bibnamefont
  {Meyer}},\ and\ \bibinfo {author} {\bibfnamefont {O.}~\bibnamefont
  {Vendrell}},\ }\bibfield  {title} {\enquote {\bibinfo {title} {Optimal mode
  combination in the multiconfiguration time-dependent hartree method through
  multivariate statistics: Factor analysis and hierarchical clustering},}\
  }\href@noop {} {\bibfield  {journal} {\bibinfo  {journal} {Journal of
  Chemical Theory and Computation}\ }\textbf {\bibinfo {volume} {19}},\
  \bibinfo {pages} {1144--1156} (\bibinfo {year} {2023})}\BibitemShut {NoStop}%
\bibitem [{\citenamefont {Schulze}\ and\ \citenamefont
  {Kuhn}(2015)}]{schulze2015explicit}%
  \BibitemOpen
  \bibfield  {author} {\bibinfo {author} {\bibfnamefont {J.}~\bibnamefont
  {Schulze}}\ and\ \bibinfo {author} {\bibfnamefont {O.}~\bibnamefont {Kuhn}},\
  }\bibfield  {title} {\enquote {\bibinfo {title} {Explicit correlated
  exciton-vibrational dynamics of the fmo complex},}\ }\href@noop {} {\bibfield
   {journal} {\bibinfo  {journal} {The Journal of Physical Chemistry B}\
  }\textbf {\bibinfo {volume} {119}},\ \bibinfo {pages} {6211--6216} (\bibinfo
  {year} {2015})}\BibitemShut {NoStop}%
\bibitem [{\citenamefont {Ishizaki}\ and\ \citenamefont
  {Fleming}(2009)}]{ishizaki2009theoretical}%
  \BibitemOpen
  \bibfield  {author} {\bibinfo {author} {\bibfnamefont {A.}~\bibnamefont
  {Ishizaki}}\ and\ \bibinfo {author} {\bibfnamefont {G.~R.}\ \bibnamefont
  {Fleming}},\ }\bibfield  {title} {\enquote {\bibinfo {title} {Theoretical
  examination of quantum coherence in a photosynthetic system at physiological
  temperature},}\ }\href@noop {} {\bibfield  {journal} {\bibinfo  {journal}
  {Proceedings of the National Academy of Sciences}\ }\textbf {\bibinfo
  {volume} {106}},\ \bibinfo {pages} {17255--17260} (\bibinfo {year}
  {2009})}\BibitemShut {NoStop}%
\bibitem [{\citenamefont {Larsson}(2019)}]{larsson_ttns_2019}%
  \BibitemOpen
  \bibfield  {author} {\bibinfo {author} {\bibfnamefont {H.~R.}\ \bibnamefont
  {Larsson}},\ }\bibfield  {title} {\enquote {\bibinfo {title} {Computing
  vibrational eigenstates with tree tensor network states (ttns)},}\ }\href
  {https://doi.org/10.1063/1.5130390} {\bibfield  {journal} {\bibinfo
  {journal} {The Journal of Chemical Physics}\ }\textbf {\bibinfo {volume}
  {151}},\ \bibinfo {pages} {204102} (\bibinfo {year} {2019})}\BibitemShut
  {NoStop}%
\bibitem [{\citenamefont {Mendive-Tapia}\ \emph {et~al.}(2017)\citenamefont
  {Mendive-Tapia}, \citenamefont {Firmino}, \citenamefont {Meyer},\ and\
  \citenamefont {Gatti}}]{mendive2017towards}%
  \BibitemOpen
  \bibfield  {author} {\bibinfo {author} {\bibfnamefont {D.}~\bibnamefont
  {Mendive-Tapia}}, \bibinfo {author} {\bibfnamefont {T.}~\bibnamefont
  {Firmino}}, \bibinfo {author} {\bibfnamefont {H.-D.}\ \bibnamefont {Meyer}},\
  and\ \bibinfo {author} {\bibfnamefont {F.}~\bibnamefont {Gatti}},\ }\bibfield
   {title} {\enquote {\bibinfo {title} {Towards a systematic convergence of
  multi-layer (ml) multi-configuration time-dependent hartree nuclear
  wavefunctions: The ml-spawning algorithm},}\ }\href@noop {} {\bibfield
  {journal} {\bibinfo  {journal} {Chemical Physics}\ }\textbf {\bibinfo
  {volume} {482}},\ \bibinfo {pages} {113--123} (\bibinfo {year}
  {2017})}\BibitemShut {NoStop}%
\end{thebibliography}%

\end{document}